%% file: 0418.tex
\documentstyle[12pt]{article}

\newcommand{\prepr}[1] {\begin{flushright}  {\bf #1} \end{flushright}
\vskip 1.cm}
\newcommand{\titul}[1] {\begin{center}{\Large {\bf #1 } } \end{center}
\vskip 0.8cm}

\newcommand{\autor}[1] {\begin{center}  {\bf \lineskip .3cm #1  }
                        \end{center} }

\newcommand{\lugar}[1] {\begin{center}  {\normalsize \bf \it #1   }
\end{center}}
\input sanda.tex

\newcounter{muni}

\begin{document}
\hbadness=10000
\pagenumbering{arabic}
\begin{titlepage}
\prepr{
Preprint hep-ph/0004xxx \\
\hspace{20mm} NCKU-HEP-00-01 \\
\hspace{20mm} APCTP-00-05 \\
\hspace{20mm} DPNU-00-14}
\titul{\bf Penguin Enhancement and
$B \rightarrow K\pi$ decays \\ in perturbative QCD}
\autor{Yong-Yeon Keum$^{1,2}$
\footnote{Email: keum@phys.sinica.edu.tw}, Hsiang-nan Li$^{2,3}$
\footnote{Email: hnli@mail.ncku.edu.tw} and 
A.I. Sanda$^4$\footnote{Email: sanda@eken.phys.nagoya-u.ac.jp}
} 
\lugar{ $^{1}$ Institute of Physics, 
Academia Sinica, Taipei, Taiwan}
\lugar{ $^{2}$ Theory Group, KEK, Tsukuba, Ibaraki 305-0801, Japan }
\lugar{ $^{3}$ Department of Physics, National Cheng-Kung University,\\
Tainan, Taiwan 701, Republic of China}
\lugar{ $^{4}$ Department of Physics, Nagoya University, Nagoya 464-01, 
Japan}

\vskip3.0cm
{\bf  PACS index : 13.25.Hw, 11.10.Hi, 12.38.Bx, 13.25.Ft}


\thispagestyle{empty}
\newpage
\vspace{10mm}
\begin{abstract}
\noindent{
We compute branching ratios of $B\to K\pi$ decays in the framework of
perturbative QCD factorization theorem. Decay amplitudes are classified
into the topologies of tree, penguin and annihilation, all of which
contain both factorizable and nonfactorizable contributions. These
contributions are expressed as the convolutions of hard $b$ quark decay
amplitudes with universal meson wave functions. It is shown that
(1) matrix elements of penguin operators are dynamically enhanced
compared to those employed in the factorization assumption; (2)
annihilation diagrams are not negligible, contrary to common belief;
(3) annihilation diagrams contribute large strong phases; (4)
the uncertainty of current data of the ratio
$R={\rm Br}(B_d^0\to K^\pm\pi^\mp)/{\rm Br}(B^\pm\to K^0\pi^\pm)$ and of
CP asymmetries is too large to give a constraint of $\phi_3$. Assuming
$\phi_3=90^o$ which is extracted from the best fit to the data of $R$,
predictions for the branching ratios of the four $B\to K\pi$ modes are
consistent with data.
}
\end{abstract}
\thispagestyle{empty}
\end{titlepage}

\newpage

\section{INTRODUCTION}

$B$ fatories at KEK and SLAC are taking data to probe the origin of CP
violation. Within the Kobayashi-Maskawa (KM) ansatz \cite{KoMa}, 
CP violation is organized in the form of a unitarity 
triangle shown in Fig.~1. The angle
$\phi_1$ can be extracted from the CP asymmetry in the $B\to J/\psi K_S$
decays, which arises from the $B$-$\bar B$ mixing. Due to similar
mechanism of CP asymmetry, the decays $B^0\to \pi^+\pi^-$ are 
appropriate for the extraction of the angle $\phi_2$. However, these 
modes contain penguin contributions such that the extraction suffers 
large uncertainty. Additional measurements of the decays 
$B^\pm\to \pi^\pm\pi^0$ and $B^0\to \pi^0\pi^0$ and the use of isospin
symmetry may resolve the uncertainties \cite{GL}. It has been proposed 
that the angle $\phi_3$ can be determined from the decays $B\to K\pi$,
$\pi\pi$ \cite{GRL,FM,NR,BF}. Contributions to these modes involve 
inteference between penguin and tree amplitudes, and relevant  
strong phases have been formulated in terms of several independent 
parameters. We are convinced that one can not make any progress
along this direction, unless he learns to compute nonleptonic
two-body decay amplitudes including strong phases.

The conventional approach to exclusive nonleptonic $B$ meson decays
relies on the factorization assumption (FA) \cite{BSW}, in which
nonfactorizable and annihilation contributions are neglected and
final-state-interaction (FSI) effects are assumed to be absent. That
is, this approach requires simplifying assumptions. 
Though analyses are easier under this assumption, estimations of 
many important ingredients, such as tree and penguin (including 
electroweak penguin) contributions, and strong phases are not reliable. 
Moreover, it suffers the problem of scale dependence \cite{CLY}.
It is also difficult to resolve some
controversies in the FA approach, such as
the branching ratios of the $B\to J/\psi K^{(*)}$ decays \cite{YL}.

Recently, perturbative QCD (PQCD) factorization theorem for exclusive
heavy-meson decays has been proved, and applied to the semileptonic
$B\to D^{(*)}(\pi)l{\bar \nu}$ decays \cite{LY1,L1}, the nonleptonic
$B\to D^{(*)}\pi(\rho)$ decays \cite{YL,CL} and the penguin-induced
radiative $B\to K^*\gamma$ decays \cite{LL,LM}. PQCD is a method to
separate hard components from a QCD process, which can be treated by
perturbation theory. Nonperturbative components are organized in the
form of hadron wave functions, which can be extracted from experimental
data. Here we shall extend the PQCD
formalism to more challenging charmless decays such as $B\to K\pi$,
$\pi\pi$. It will be shown that the difficulties encountered in the
FA approach can be resolved in the PQCD formalism.

In this paper we shall evaluate the branching ratios of the following
modes,
\begin{eqnarray}
& &B^\pm\to K^0\pi^\pm\;,\;\;\;\;B_d^0\to K^\pm\pi^\mp\;, 
\nonumber\\
& &B^\pm\to K^\pm\pi^0\;,\;\;\;\;B_d^0\to K^0\pi^0\;.
\label{bkpi}
\end{eqnarray}
Contributions from various topologies, such as tree, penguin and
annihilation, including both factorizable and nonfactorizable
contributions, can all be calculated. That is,  
FA is in fact not necessary. In our approach strong phases arise
from non-pinched singularities of quark and gluon propagators in
nonfactorizable and annihilation diagrams. As explicitly shown in
Sec.~VII, strong phases from the Bander-Silverman-Soni (BSS) mechanism
\cite{BSS}, which is a source of strong phases in the FA
approach, are of next-to-leading order and negligible.

As an application, we derive the ratio $R$ and the CP asymmetries
defined by
\begin{eqnarray}
R&=&\frac{{\rm Br}(B_d^0\to K^\pm\pi^\mp)}{{\rm Br}
(B^\pm\to K^0\pi^\pm)}\;,
\label{tr}
\\
A_{CP}^0&=&\frac{{\rm Br}({\bar B}_d^0\to K^-\pi^+)
- {\rm Br}(B_d^0\to K^+\pi^-)}
{{\rm Br}({\bar B}_d^0\to K^-\pi^+) + {\rm Br}(B_d^0\to K^+\pi^-)}\;,
\label{cp1}\\
A_{CP}^c&=&\frac{{\rm Br}(B^-\to K^0\pi^-)-{\rm Br}(B^+\to K^0\pi^+)}
{{\rm Br}(B^-\to K^0\pi^-)+{\rm Br}(B^+\to K^0\pi^+)}\;,
\\
A_{CP}^{'0}&=&\frac{{\rm Br}({\bar B}_d^0\to K^0\pi^0)
- {\rm Br}(B_d^0\to {\bar K}^0\pi^0)}
{{\rm Br}({\bar B}_d^0\to K^0\pi^0)
+ {\rm Br}(B_d^0\to {\bar K}^0\pi^0)}\;,
\\
A_{CP}^{'c}&=&\frac{{\rm Br}(B^-\to K^-\pi^0)-{\rm Br}(B^+\to K^+\pi^0)}
{{\rm Br}(B^-\to K^-\pi^0)+{\rm Br}(B^+\to K^+\pi^0)}\;,
\label{cp}
\end{eqnarray}
as functions of the unitarity angle $\phi_3$ using PQCD factorization 
theorem. In the above expressions ${\rm Br}(B_d^0\to K^\pm\pi^\mp)$
represents the CP average of the branching ratios
${\rm Br}(B_d^0\to K^+\pi^-)$ and ${\rm Br}({\bar B}_d^0\to K^-\pi^+)$,
and the definition of ${\rm Br}(B^\pm\to K^0\pi^\pm)$ is similar. 
It will be shown that the uncertainty in the data for $R$, $A_{CP}^0$ and
$A_{CP}^c$ \cite{YK,CLEO3},
\begin{equation}
R=0.95\pm 0.30\;,\;\;\;\;
A_{CP}^0=-0.04\pm 0.16\;,\;\;\;\;
A_{CP}^c=0.17\pm 0.24\;,
\label{cld}
\end{equation}
is still too large to provide useful information of $\phi_3$.
Using the central values of the CLEO data for $R$,
we obtain $\phi_3= 90^o$.

An essential difference between the FA and PQCD approaches is
that the hard scale at which Wilson coefficients are evaluated is chosen
arbitrarily as $m_b$ or $m_b/2$ in the former, $m_b$ being the $b$ quark
mass, but dynamically determined in the latter. It has been shown that
choosing this dynamically determined scale
minimizes higher-order corrections to exclusive QCD processes
\cite{MN}. We observe that this choice leads to an
enhancement of penguin contributions by nearly 50\% compared to those
in the FA approach. As elaborated in Sec.~V, this penguin
enhancement is crucial for the explanation of the data of all $B\to K\pi$,
$\pi\pi$ modes using a smaller angle $\phi_3\sim 90^o$. Note that an
angle $\phi_3$ larger than $110^o$ must be adopted in order to explain
the above data in the FA approach \cite{WS}.

PQCD factorization theorem for exclusive nonleptonic $B$ meson decays
are briefly reviewed in Sec.~II. The factorization formulas for various
$B\to K\pi$ decay modes are derived in Sec.~III. The numerical analysis,
including the determination of meson wave functions, is performed in
Sec.~IV. We emphasize the importance of the penguin enhancement in the
PQCD approach in Sec.~V. FSI effects are discussed in Sec.~VI. The PQCD
approach is compared with other approaches in Sec.~VII.
Section VIII is the conclusion.

\section{FACTORIZATION THEOREM IN BRIEF}

We first sketch the rough idea of PQCD factorization theorem and of its
application to two-body $B$ meson decays.
Take the $B\to\pi$ transition form factor in the fast recoil region of
the pion as an example. Obviously, this process involves two scales: the
$b$ quark mass $m_b$, which provides the large energy release to the fast
pion, and the QCD scale $\Lambda_{\rm QCD}$, which is associated with
bound-state mesons. Therefore, the $B\to\pi$ transition form factor
contains both perturbative and nonperturbative dynamics.

In perturbation theory nonperturbative dynamics is reflected by infrared
divergences in radiative corrections. It has been shown order by order
that these infrared divergences can be separated and absorbed into a $B$
meson wave function or a pion wave function \cite{LY1}. A formal
definition of the
meson wave functions as matrix elements of nonlocal operators can be
constructed, which, if evaluated perturbatively, reproduces the infrared
divergences. Certainly, one can not derive a wave function using a
perturbative method, but has to parametrize it as a parton model, which
describes how a parton (valence quark, if a leading-twist wave function
is referred) shares meson momentum. The meson wave functions,
characterized by $\Lambda_{\rm QCD}$, must be determined by
nonperturbative means, such as lattice gauge theory and QCD sum rules, or
extracted from experimental data. In the practical calculation below,
small parton transverse momenta $k_T$ are included, and the
characteristic scale is replaced by $1/b$ with $b$ being a variable
conjugate to $k_T$.

After absorbing infrared divergences into the meson wave functions, the
remaining part of radiative corrections is infrared finite. This part,
called a hard amplitude, can be evaluated perturbatively in terms of
Feynman diagrams for decays of an on-shell $b$ quark. Note that
the $b$ quark carries various momenta, whose distribution is described
by the parton model introduced above. The hard decay amplitude is
characterized by the virtuality $t$ of involved internal particles, which
is of order $m_b$. The $B\to\pi$ transition form factor is then expressed
as the convolution of three factors: the $B$ meson and pion wave
functions, and the hard $b$ quark decay amplitude. This is so called
factorization theorem. Note that the separation of nonperturbative and
perturbative dynamics is quite arbitrary. This arbitrariness implies
that a renormalization-group (RG) improvement of the factorization
formula for the $B\to\pi$ transition form factor can be implemented. The
RG evolution from the all-order summation of large logarithmic
corrections to the above convolution factors will be made explicit below.

A salient feature of PQCD factorization theorem is the universality
of nonperturbative wave functions. Briefly speaking, the infrared
divergences associated with the $B$ meson are process-independent,
and the formal definition of the $B$ meson wave function in terms of
matrix elements of nonlocal operators is universal for
all $B$ meson decay modes. It is not difficult to understand this
universality: infrared divergences correspond to
long-distance effects, while the hard $b$ quark decay occurs in a very
short space-time. It is natural that these two dramastically
different subprocesses decouple from each other. That is, the
long-distance dynamics is insensitive to specific decays of the $b$
quark with large energy release. Because of
universality, a $B$ meson wave function extracted
from some decay modes can be employed to make predictions for other
modes. This is the reason PQCD factorization theorem possesses a
predictive power. We emphasize that PQCD is a theory, instead of a model,
since higher-order and higher-twist contributions can be included
systematically. The model independence of PQCD predictions can be
achieved, once wave functions are determined precisely.

PQCD factorization theorem for nonleptonic $B$ meson decays, such as
$B\to K(\pi)\pi$ and $B\to D^{(*)}\pi(\rho)$, is more complicated.
These decays involve three scales: the $W$ boson mass $M_W$, at which
the matching conditions of the effective weak Hamiltonian to the full
Hamiltonian are defined, the typical scale $t$ (of order $m_b$) of a
hard amplitude, which reflects the specific dynamics of a decay mode,
and the factorization scale $1/b$ (of order $\Lambda_{\rm QCD}$)
introduced above. The dynamics below $1/b$ is regarded as being
completely nonperturbative, and parametrized into meson wave functions
$\phi(x,b)$, $x$ being the momentum fraction. Above the factorization
scale, radiative corrections to on-shell $b$ quark decays can be
computed perturbatively. This part contains two
characteristic scales, $M_W$ and $t$, differing from the case of
the $B\to \pi$ transition from factor, and further factorization is
necessary.

Radiative corrections produce two types of large logarithms:
$\ln(M_W/t)$ and $\ln(tb)$. The former are summed by RG equations to give
the evolution from $M_W$ down to $t$ described by the Wilson coefficients
$C(t)$, while the latter are summed to give the evolution from $t$ to
$1/b$. The matching between the full Hamiltonian and the effective
Hamiltonian in the above three-scale factorization theorem is similar to
that in the standard effective field theory. The difference is that
diagrams in the full theory contain not only $W$ boson emissions but hard
gluon emissions from spectator quarks \cite{CLY}. One can show that the
effective operators, in the presence of the hard gluons from spectators,
still form a complete basis, and that the Wilson coefficients derived in
the three-scale factorization theorem are the same as those derived in
the standard effective theory.

There also exist double logarithms $\ln^2(Pb)$ from the overlap of two
types of infrared divergences, collinear and soft, in radiative
corrections to meson wave functions \cite{CS}, where $P$ denotes the
dominant light-cone component of a meson momentum. The resummation
\cite{CS,BS} of these double logarithms leads to a Sudakov form factor
$\exp[-s(P,b)]$, which suppresses the long-distance contributions from
the large $b$ region, and vanishes as $b=1/\Lambda_{\rm QCD}$. This factor
guarantees the applicability of PQCD to exclusive decays around the energy 
scale of the $b$ quark mass \cite{LY1}. For a detailed derivation of the 
relevant Sudakov form factors, we refer the readers to \cite{LY1,L1}.
With all the large logarithms organized, the remaining finite contributions 
are absorbed into the hard $b$ quark decay amplitude $H(t)$.

A three-scale factorization formula for exclusive nonleptonic $B$ meson
decays possesses the typical expression,
\begin{eqnarray}
C(t)\otimes H(t)\otimes \phi(x,b)
\otimes\exp\left[-s(P,b)-2\int_{1/b}^t\frac{d{\bar\mu}}
{\bar\mu}\gamma(\alpha_s({\bar\mu}))\right],
\label{for}
\end{eqnarray}
where the exponential involving the quark anomalous dimension
$\gamma=-\alpha_s/\pi$ describes the evolution from $t$ to $1/b$
mentioned above. Note that Eq.~(\ref{for}) is a convolution relation,
with internal parton kinematics $x$ and $b$ integrated out. 
The hard scale $t$, related to the virtuality of internal particles
in hard amplitudes, depends on $x$ and $b$. All the convolution factors,
except for the wave functions $\phi(x,b)$, are calculable in perturbation
theory. The wave functions, though not calculable, are universal. If
choosing $t$ as the $b$ quark mass $m_b$, the Wilson coefficient $C(m_b)$
is a constant, and Eq.~(\ref{for}) reduces to the simple product of the
Wilson coefficient and a hadronic matrix element. However, the analysis 
of next-to-leading-order corrections to the pion form factor \cite{MN} has 
suggested that $t$ should be chosen as the virtuality of internal 
particles in order to minimize higher-order corrections to the hard
amplitudes.

Because of the universality of nonperturbative wave functions stated 
above, the strategy of PQCD factorization theorems is as follows:
evaluate all perturbative factors for some decay modes, and adjust the
wave functions such that predictions from the corresponding factorization
formulas match experimental data. At this stage, the nonperturbative wave
functions are determined up to the twist, at which the factorization 
is constructed. Then
evaluate all the perturbative factors for another decay mode. Input the
extracted wave functions into the factorization formulas of the same
twist, and make predictions. With this strategy, PQCD
factorization theorems are model-independent and possess a predictive
power. In Sec.~IV we shall make explicit the determination of the
$B$ meson, kaon, and pion wave functions from currently available data
and phenomenological arguments.

\section{$B\to K\pi$ AMPLITUDES}

The effective Hamiltonian for the flavor-changing $b\to s$ transition is
given by \cite{REVIEW}
\begin{equation}
H_{\rm eff}={G_F\over\sqrt{2}}
\sum_{q=u,c}V_q\left[C_1(\mu)O_1^{(q)}(\mu)+C_2(\mu)O_2^{(q)}(\mu)+
\sum_{i=3}^{10}C_i(\mu)O_i(\mu)\right]\;,
\label{hbk}
\end{equation}
with the Cabibbo-Kobayashi-Maskawa (CKM)
matrix elements $V_q=V^*_{qs}V_{qb}$ and the operators
\begin{eqnarray}
& &O_1^{(q)} = (\bar{s}_iq_j)_{V-A}(\bar{q}_jb_i)_{V-A}\;,\;\;\;\;\;\;\;\;
O_2^{(q)} = (\bar{s}_iq_i)_{V-A}(\bar{q}_jb_j)_{V-A}\;, 
\nonumber \\
& &O_3 =(\bar{s}_ib_i)_{V-A}\sum_{q}(\bar{q}_jq_j)_{V-A}\;,\;\;\;\;
O_4 =(\bar{s}_ib_j)_{V-A}\sum_{q}(\bar{q}_jq_i)_{V-A}\;, 
\nonumber \\
& &O_5 =(\bar{s}_ib_i)_{V-A}\sum_{q}(\bar{q}_jq_j)_{V+A}\;,\;\;\;\;
O_6 =(\bar{s}_ib_j)_{V-A}\sum_{q}(\bar{q}_jq_i)_{V+A}\;, 
\nonumber \\
& &O_7 =\frac{3}{2}(\bar{s}_ib_i)_{V-A}\sum_{q}e_q(\bar{q}_jq_j)_{V+A}\;,
\;\;
O_8 =\frac{3}{2}(\bar{s}_ib_j)_{V-A}\sum_{q}e_q(\bar{q}_jq_i)_{V+A}\;, 
\nonumber \\
& &O_9 =\frac{3}{2}(\bar{s}_ib_i)_{V-A}\sum_{q}e_q(\bar{q}_jq_j)_{V-A}\;,
\;\;
O_{10} =\frac{3}{2}(\bar{s}_ib_j)_{V-A}\sum_{q}e_q(\bar{q}_jq_i)_{V-A}\;, 
\end{eqnarray} 
$i, \ j$ being the color indices. Using the unitarity condition, the 
CKM matrix elements for the penguin operators $O_3$-$O_{10}$ can also
be expressed as $V_u+V_c=-V_t$. 
We define the angle $\phi_3$ via
\begin{equation}
V_{ub}=|V_{ub}|\exp(-i\phi_3)\;.
\end{equation} 
Here we adopt the Wolfstein parametrization for the
CKM matrix upto ${\cal O}(\lambda^{3})$:
\begin{eqnarray}
\left(\matrix{V_{ud} & V_{us} & V_{ub} \cr
              V_{cd} & V_{cs} & V_{cb} \cr
              V_{td} & V_{ts} & V_{tb} \cr}\right)
=\left(\matrix{ 1 - { \lambda^2 \over 2 } & \lambda &
A \lambda^3(\rho - i \eta)\cr
- \lambda & 1 - { \lambda^2 \over 2 } & A \lambda^2\cr
A \lambda^3(1-\rho-i\eta) & -A \lambda^2 & 1 \cr}\right)\;.
\end{eqnarray} 
A recent analysis of quark-mixing matrix yields \cite{LEP}
\begin{eqnarray}
\lambda &=& 0.2196 \pm 0.0023\;,
\nonumber \\
A &=& 0.819 \pm 0.035\;,
\nonumber \\
R_b &\equiv&\sqrt{{\rho}^2  + {\eta}^2}
= 0.41 \pm 0.07\;. 
\end{eqnarray}

For the $B^\pm\to K^0\pi^\pm$ decays, the operators $O_{1,2}^{(u)}$ 
contribute via an annihilation topology, and $O_{1,2}^{(c)}$ do not
contribute at leading order of $\alpha_s$. The absorptive part of the
charm quark loop integral computed by BSS is thus
of higher order. $O_{3,4,5,6}$ contribute via
tree and annihilation topologies, and the tree topology involves 
the $B\to \pi$ form factor. $O_{3,5}$ gives both factorizable and
nonfactorizable (color-suppressed) contributions, while $O_{4,6}$ gives
only factorizable ones because of the color flow. The contributions from
$O_{7,8,9,10}$ are the same as $O_{3,4,5,6}$ except for an additional
factor $(3/2)e_q$ with the light quark $q=d$ in the tree topology and
with $q=u$ in the annihilation topology. For the $B_d^0\to K^\pm\pi^\mp$
decays, the operators $O_{1,2}^{(u)}$ contribute via a tree topology, and
$O_{1,2}^{(c)}$ do not contribute at leading order of $\alpha_s$. The
penguin operators contribute in the same way as in the
$B^\pm\to K^0\pi^\pm$ decays but with the light quark $q=u$ in the tree
topology and with $q=d$ in the annihilation topology. The lowest-order
hard $b$ quark decay amplitudes are summarized in Fig.~2
for $B_d^{0} \to K^{\mp}\pi^{\pm}$ decays 
and in Fig.~3 for $B^{\pm} \to \bar{K}^{0}\pi^{\pm}$ decays.

For the $B^\pm\to K^\pm\pi^0$ decays, the operators $O_{1,2}^{(u)}$ 
contribute via tree and annihilation topologies, where the tree topology
involves both the $B\to \pi$ and $B\to K$ form factors. The penguin
operators also contribute via tree and annihilation topologies 
with the light quark $q=u$ in both topologies. While the tree topology
involves only the $B\to\pi$ form factor. For the $B_d^0\to K^0\pi^0$
decays, the operators $O_{1,2}^{(u)}$ contribute via the tree topology,
which involves only the $B\to K$ form factor. The penguin operators
contribute in a similar way but with the light quark $q=d$ in both
the tree and annihilation topologies. Their lowest-order diagrams
for the hard $b$ quark decay amplitudes are basically similar to those
in Figs.~2 and 3.

The momenta of the $B$ and $K$ mesons in light-cone coordinates are written
as $P_1=(M_B/\sqrt{2})(1,1,{\bf 0}_T)$ and 
$P_2=(M_B/\sqrt{2})(1,0,{\bf 0}_T)$, respectively.
The $B$ meson is at rest with the
above parametrization of momenta. We define the momenta of light valence
quark in the $B$ meson as $k_1$, where $k_1$ has a plus component $k_1^+$,
giving the momentum fraction $x_1=k_1^+/P_1^+$, and small
transverse components ${\bf k}_{1T}$. The light valence quark and the
$s$ quark in the kaon carry the longitudinal momenta $x_2P_2$ and
$(1-x_2)P_2$, and small transverse momenta ${\bf k}_{2T}$ and
$-{\bf k}_{2T}$, respectively. The
pion momentum is then $P_3=P_1-P_2$, whose nonvanishing component is only
$P_3^-$. The two light valence quarks in the pion carry the longitudinal 
momenta $x_3P_3$ and $(1-x_3)P_3$, and small transverse momenta 
${\bf k}_{3T}$ and $-{\bf k}_{3T}$, respectively. The kinematic
variables associated with each meson are indicated in Fig.~4.

The Sudakov resummations of the large logarithmic corrections to the $B$,
$K$ and $\pi$ meson wave functions $\phi_B$, $\phi_K$ and $\phi_\pi$
lead to the exponentials $\exp(-S_B)$, $\exp(-S_{K})$ and $\exp(-S_\pi)$,
respectively, with the exponents \cite{LY1,LS}
\begin{eqnarray}
S_B(t)&=&s(x_1P_1^+,b_1)+2\int_{1/b_1}^{t}\frac{d{\bar \mu}}{\bar \mu}
\gamma(\alpha_s({\bar \mu}))\;,
\nonumber \\
S_{K}(t)&=&s(x_2P_2^+,b_2)+s((1-x_2)P_2^+,b_2)+2\int_{1/b_2}^{t}
\frac{d{\bar \mu}}{\bar \mu}\gamma(\alpha_s({\bar \mu}))\;,
\nonumber\\
S_\pi(t)&=&s(x_3P_3^-,b_3)+s((1-x_3)P_3^-,b_3)+2\int_{1/b_3}^{t}
\frac{d{\bar \mu}}{\bar \mu}\gamma(\alpha_s({\bar \mu}))\;.
\label{sbk}
\end{eqnarray}
The variables $b_1$, $b_2$, and $b_3$ conjugate to the parton transverse 
momentum $k_{1T}$, $k_{2T}$, and $k_{3T}$ represent the transverse 
extents of the $B$, $K$, and $\pi$ meson, respectively. The exponent $s$ 
is written as \cite{CS,BS}
\begin{equation}
s(Q,b)=\int_{1/b}^{Q}\frac{d \mu}{\mu}\left[\ln\left(\frac{Q}{\mu}
\right)A(\alpha_s(\mu))+B(\alpha_s(\mu))\right]\;,
\label{fsl}
\end{equation}
where the anomalous dimensions $A$ to two loops and $B$ to one loop are
\begin{eqnarray}
A&=&C_F\frac{\alpha_s}{\pi}+\left[\frac{67}{9}-\frac{\pi^2}{3}
-\frac{10}{27}f+\frac{2}{3}\beta_0\ln\left(\frac{e^{\gamma_E}}{2}\right)
\right]\left(\frac{\alpha_s}{\pi}\right)^2\;,
\nonumber \\
B&=&\frac{2}{3}\frac{\alpha_s}{\pi}\ln\left(\frac{e^{2\gamma_E-1}}
{2}\right)\;,
\end{eqnarray}
with $C_F=4/3$ a color factor, $f=4$ the active flavor number, and 
$\gamma_E$ the Euler constant.
The one-loop expression of the running coupling constant,
\begin{equation}
\alpha_s(\mu)=\frac{4\pi}{\beta_0\ln(\mu^2/\Lambda^2)}\;,
\end{equation}
is substituted into Eq.~(\ref{fsl}) with the coefficient
$\beta_{0}=(33-2f)/3$.

The decay rates of $B^\pm\to K^0\pi^\pm$ have the expressions
\begin{equation}
\Gamma=\frac{G_F^2M_B^3}{128\pi}|{\cal A}|^2\;.
\label{dr1}
\end{equation}
The decay amplitudes ${\cal A}^+$ and ${\cal A}^-$ corresponding to
$B^+\to K^0\pi^+$ and $B^-\to K^0\pi^-$, respectively, are written as
\begin{eqnarray}
{\cal A^+}&=&f_KV_t^*F^P_e+V_t^*{\cal M}^P_e
+f_BV_t^*F^P_a+V_t^*{\cal M}^P_a
-f_BV_u^*F_a-V_u^*{\cal M}_a\;,
\label{Map}\\
{\cal A^-}&=&f_KV_tF^P_e+V_t{\cal M}^P_e
+f_BV_tF^P_a+V_t{\cal M}^P_a
-f_BV_uF_a-V_u{\cal M}_a\;,
\label{Mam}
\end{eqnarray}
with the $B$ meson (kaon) decay constant $f_{B(K)}$. The notations $F$
represent factorizable contributions (form factors), and ${\cal M}$
represent nonfactorizable (color-suppressed) contributions. The indices
$a$, $e$ and $P$ denote the annihilation, tree and penguin topologies,
respectively. $F_a$, associated with the time-like $K\to\pi$ form
factor, and ${\cal M}_a$ are from the operators $O^{(u)}_{1,2}$.
The decay rates of $B_d^0\to K^\pm\pi^\mp$ have the similar expressions
with amplitudes
\begin{eqnarray}
{\cal A}&=&f_KV_t^*F^P_e+V_t^*{\cal M}^P_e
+f_BV_t^*F^P_a+V_t^*{\cal M}^P_a
-f_KV_u^*F_e-V_u^*{\cal M}_e\;,
\label{Mbp}\\
{\bar{\cal A}}&=&f_KV_tF^P_e
+V_t{\cal M}^P_e
+f_BV_tF^P_a+V_t{\cal M}^P_a
-f_KV_uF_e-V_u{\cal M}_e\;,
\label{Mbm}
\end{eqnarray}
for $B_d^0\to K^+\pi^-$ and ${\bar B}_d^0\to K^-\pi^+$, respectively.
The notations are similar to those in Eqs.~(\ref{Map}) and (\ref{Mam}).
$F_e$, associated with the $B\to\pi$ form factor, and ${\cal M}_e$ are
from the operators $O^{(u)}_{1,2}$.

The decay amplitudes for $B^\pm\to K^\pm\pi^0$ are given by
\begin{eqnarray}
\sqrt{2}{\cal A}^{'+}&=&f_KV_t^*F^P_e+V_t^*{\cal M}^P_e
+f_BV_t^*F^P_a+V_t^*{\cal M}^P_a
-f_BV_u^*F_a-V_u^*{\cal M}_a
\nonumber\\
& &-f_KV_u^*F_e-V_u^*{\cal M}_e-f_\pi V_u^*F_{eK}-V_u^*{\cal M}_{eK}\;,
\label{Mapp}\\
\sqrt{2}{\cal A}^{'-}&=&f_KV_tF^P_e+V_t{\cal M}^P_e
+f_BV_tF^P_a+V_t{\cal M}^P_a
-f_BV_uF_a-V_u{\cal M}_a
\nonumber\\
& &-f_KV_uF_e-V_u{\cal M}_e-f_\pi V_uF_{eK}-V_u{\cal M}_{eK}\;,
\label{Mamp}
\end{eqnarray}
which correspond to $B^+\to K^+\pi^0$ and $B^-\to K^-\pi^0$, 
respectively. The factorizable contribution $F_{eK}$ is associated 
with the $B\to K$ form factor, and
${\cal M}_{eK}$ is the corresponding nonfactorizable contribution.
Similarly, the decay rates of $B_d^0\to K^0\pi^0$ are obtained
from the amplitudes
\begin{eqnarray}
\sqrt{2}{\cal A}'&=&f_KV_t^*F^P_e+V_t^*{\cal M}^P_e
+f_BV_t^*F^P_a+V_t^*{\cal M}^P_a
-f_\pi V_u^*F_{eK}-V_u^*{\cal M}_{eK}\;,
\label{Mbpp}\\
\sqrt{2}{\bar{\cal A}}'&=&f_KV_tF^P_e+V_t{\cal M}^P_e
+f_BV_tF^P_a+V_t{\cal M}^P_a
-f_\pi V_uF_{eK}-V_u{\cal M}_{eK}\;,
\label{Mbmp}
\end{eqnarray}
for $B_d^0\to K^0\pi^0$ and ${\bar B}_d^0\to {\bar K}^0\pi^0$, 
respectively.

Basically, one needs to derive the factorizaton formulas only for the
tree and annihilation topologies. Wilson coefficients corresponding to
different operators are then inserted into the factorization formulas.
The form factors are written as
\begin{eqnarray}
F^P_e&=&F^P_{e4}+F^P_{e6}\;,
\nonumber \\
F^P_{e4}&=&16\pi C_FM_B^2\int_0^1 dx_1dx_3\int_0^{\infty}b_1db_1b_3db_3
\phi_B(x_1,b_1)
\nonumber \\
& &\times\{\left[(1+x_3)\phi_\pi(x_3)+r_\pi(1-2x_3)\phi'_\pi(x_3)
\right]E_{e4}(t^{(1)}_e)h_e(x_1,x_3,b_1,b_3)
\nonumber\\
& &+2r_\pi\phi'_\pi(x_3)E_{e4}(t^{(2)}_e)
h_e(x_3,x_1,b_3,b_1)\}\;,
\label{int4}\\
F^P_{e6}&=&32\pi C_FM_B^2\int_0^1 dx_1dx_3\int_0^{\infty}b_1db_1b_3db_3
\phi_B(x_1,b_1)
\nonumber \\
& &\times r_K\{\left[\phi_\pi(x_3)+r_\pi(2+x_3)\phi'_\pi(x_3)
\right]E_{e6}(t^{(1)}_e)h_e(x_1,x_3,b_1,b_3)
\nonumber\\
& &+\left[x_1\phi_\pi(x_3)+2r_\pi(1-x_1)\phi'_\pi(x_3)\right]
E_{e6}(t^{(2)}_e)h_e(x_3,x_1,b_3,b_1)\}\;,
\label{int6}\\
F^P_a&=&F^P_{a4}+F^P_{a6}\;,
\nonumber\\
F^P_{a4}&=&16\pi C_FM_B^2\int_0^1 dx_2dx_3\int_0^{\infty}b_2db_2b_3db_3
\nonumber \\
& &\times\{\left[-x_3\phi_{K}(x_2)\phi_\pi(x_3)
-2r_\pi r_K(1+x_3)\phi'_{K}(x_2)\phi'_\pi(x_3)\right]
E_{a4}(t^{(1)}_a)h_a(x_2,x_3,b_2,b_3)
\nonumber\\
& &+\left[x_2\phi_{K}(x_2)\phi_\pi(x_3)
+2r_\pi r_K(1+x_2)\phi'_{K}(x_2)\phi'_\pi(x_3)\right]
E_{a4}(t^{(2)}_a)h_a(x_3,x_2,b_3,b_2)\}\;,
\label{exc4}\\
F^P_{a6}&=&32\pi C_FM_B^2\int_0^1 dx_2dx_3\int_0^{\infty}b_2db_2b_3db_3
\nonumber \\
& &\times\{\left[r_\pi x_3\phi_{K}(x_2)\phi'_\pi(x_3)
+2r_K\phi'_{K}(x_2)\phi_\pi(x_3)\right]
E_{a6}(t^{(1)}_a)h_a(x_2,x_3,b_2,b_3)
\nonumber\\
& &+\left[2r_\pi\phi_{K}(x_2)\phi'_\pi(x_3)
+r_Kx_2\phi'_{K}(x_2)\phi_\pi(x_3)\right]
E_{a6}(t^{(2)}_a)h_a(x_3,x_2,b_3,b_2)\}\;,
\label{exc6}
\end{eqnarray}
with the evolution factors
\begin{eqnarray}
E_{ei}(t)&=&\alpha_s(t)a_i(t)\exp[-S_B(t)-S_\pi(t)]\;,
\\
E_{ai}(t)&=&\alpha_s(t)a_i(t)\exp[-S_K(t)-S_\pi(t)]\;.
\end{eqnarray}
The expression of $F_e$ ($F_a$) for the $O_{1,2}$ contributions 
is the same as $F^P_{e4}$ ($F^P_{a4}$)
but with the Wilson coefficient $a_1(t_e)$ ($a_1(t_a)$).
The factorization formula of $F_{eK}$ is written as
\begin{eqnarray}
F_{eK}&=&16\pi C_FM_B^2\int_0^1 dx_1dx_2\int_0^{\infty}b_1db_1b_2db_2
\phi_B(x_1,b_1)
\nonumber \\
& &\times\{\left[(1+x_2)\phi_K(x_2)+r_K(1-2x_2)\phi'_K(x_2)
\right]E_2(t^{(1)}_{eK})h_e(x_1,x_2,b_1,b_2)
\nonumber\\
& &+2r_K\phi'_K(x_2)E_2(t^{(2)}_{eK})
h_e(x_2,x_1,b_2,b_1)\}\;,
\label{intk}
\end{eqnarray}
with the evolution factor
\begin{eqnarray}
E_2(t)=\alpha_s(t)a_2(t)\exp[-S_B(t)-S_K(t)]\;.
\end{eqnarray}

The hard functions $h$'s in Eqs~(\ref{int4})-(\ref{exc6}) and in
Eq.~(\ref{intk}), are given by
\begin{eqnarray}
h_e(x_1,x_3,b_1,b_3)&=&K_{0}\left(\sqrt{x_1x_3}M_Bb_1\right)
\nonumber \\
& &\times \left[\theta(b_1-b_3)K_0\left(\sqrt{x_3}M_B
b_1\right)I_0\left(\sqrt{x_3}M_Bb_3\right)\right.
\nonumber \\
& &\left.+\theta(b_3-b_1)K_0\left(\sqrt{x_3}M_Bb_3\right)
I_0\left(\sqrt{x_3}M_Bb_1\right)\right]\;,
\label{dh}\\
h_a(x_2,x_3,b_2,b_3)&=&\left(\frac{i\pi}{2}\right)^2
H_0^{(1)}\left(\sqrt{x_2x_3}M_Bb_2\right)
\nonumber \\
& &\times\left[\theta(b_2-b_3)
H_0^{(1)}\left(\sqrt{x_3}M_Bb_2\right)
J_0\left(\sqrt{x_3}M_Bb_3\right)\right.
\nonumber \\
& &\left.+\theta(b_3-b_2)H_0^{(1)}\left(\sqrt{x_3}M_Bb_3\right)
J_0\left(\sqrt{x_3}M_Bb_2\right)\right]\;.
\label{ah}
\end{eqnarray}
The derivation of $h$, from the Fourier transformation of the
lowest-order $H$, is the same as that for the $B\to D\pi$ decays
\cite{YL}, but with a vanishing $D$ meson mass.
The hard scales $t$ are chosen as the maxima of the virtualities of
internal particles involved in $b$ quark decay amplitudes, including
$1/b_i$:
\begin{eqnarray}
t^{(1)}_e&=&{\rm max}(\sqrt{x_3}M_B,1/b_1,1/b_3)\;,
\nonumber\\
t^{(2)}_e&=&{\rm max}(\sqrt{x_1}M_B,1/b_1,1/b_3)\;,
\nonumber\\
t^{(1)}_a&=&{\rm max}(\sqrt{x_3}M_B,1/b_2,1/b_3)\;,
\nonumber\\
t^{(2)}_a&=&{\rm max}(\sqrt{x_2}M_B,1/b_2,1/b_3)\;,
\nonumber\\
t^{(1)}_{eK}&=&{\rm max}(\sqrt{x_2}M_B,1/b_1,1/b_2)\;,
\nonumber\\
t^{(2)}_{eK}&=&{\rm max}(\sqrt{x_1}M_B,1/b_1,1/b_2)\;,
\label{et}
\end{eqnarray}
which decrease higher-order corrections.
The Sudakov factor in Eq.~(\ref{sbk})
suppresses long-distance contributions from the large $b$ region, and
improves the applicability of PQCD to $B$ meson decays.

For the nonfactorizable amplitudes, the factorization formulas involve
the kinematic variables of all the three mesons, and the Sudakov exponent
is given by $S=S_B+S_K+S_\pi$. The integration over $b_3$ can be
performed trivially, leading to $b_3=b_1$, $b_3=b_2$, or $b_2=b_1$. Their
expressions are
\begin{eqnarray}
{\cal M}^P_e&=&{\cal M}^P_{e4}+{\cal M}^P_{e6}\;,
\nonumber \\
{\cal M}^P_{e4}&=&32\pi C_F\sqrt{2N_c}M_B^2\int_0^1 [dx]\int_0^{\infty}
b_1 db_1 b_2 db_2\phi_B(x_1,b_1)\phi_K(x_2)
\nonumber \\
& &\times \{\left[(x_1-x_2-x_3)\phi_\pi(x_3)
+r_\pi x_3\phi'_\pi(x_3)\right]E'_{e4}(t^{(1)}_d)
h^{(1)}_d(x_1,x_2,x_3,b_1,b_2,b_1)
\nonumber \\
& &+\left[(1-x_1-x_2)\phi_\pi(x_3)-r_\pi x_3\phi'_\pi(x_3)\right]
E'_{e4}(t^{(2)}_d)h^{(2)}_d(x_1,x_2,x_3,b_1,b_2,b_1)\}\;,
\label{md3}\\
{\cal M}^P_{e6}&=&32\pi C_F\sqrt{2N_c}M_B^2\int_0^1 [dx]\int_0^{\infty}
b_1 db_1 b_2 db_2\phi_B(x_1,b_1)\phi'_K(x_2)
\nonumber \\
& &\times r_K\{\left[(x_1-x_2)\phi_\pi(x_3)
+r_\pi (x_1-x_2-x_3)\phi'_\pi(x_3)\right]
E'_{e6}(t^{(1)}_d)h^{(1)}_d(x_1,x_2,x_3,b_1,b_2,b_1)
\nonumber \\
& &+\left[(1-x_1-x_2)\phi_\pi(x_3)+r_\pi (1-x_1-x_2+x_3)
\phi'_\pi(x_3)\right]
\nonumber\\
& &\hspace{1.0 cm}\times E'_{e6}(t^{(2)}_d)
h^{(2)}_d(x_1,x_2,x_3,b_1,b_2,b_1)\}\;,
\label{md5}\\
{\cal M}^P_a&=&{\cal M}^P_{a4}+{\cal M}^P_{a6}\;,
\nonumber \\
{\cal M}^P_{a4}&=&32\pi C_F\sqrt{2N_c}M_B^2\int_0^1 [dx]\int_0^{\infty}
b_1 db_1 b_2 db_2\phi_B(x_1,b_1)
\nonumber \\
& &\times \{\left[x_3\phi_K(x_2)\phi_\pi(x_3)
-r_\pi r_K(x_1-x_2-x_3)\phi'_K(x_2)\phi'_\pi(x_3)\right]
\nonumber \\
& &\hspace{1.0cm}
\times E'_{a4}(t^{(1)}_f)h^{(1)}_f(x_1,x_2,x_3,b_1,b_2,b_2)
\nonumber \\
& &-\left[(x_1+x_2)\phi_K(x_2)\phi_\pi(x_3)
+r_\pi r_K(2+x_1+x_2+x_3)\phi'_K(x_2)\phi'_\pi(x_3)\right]
\nonumber\\
& &\hspace{1.0cm}\times E'_{a4}(t^{(2)}_f)
h^{(2)}_f(x_1,x_2,x_3,b_1,b_2,b_2)\}\;,
\label{mf3}\\
{\cal M}^P_{a6}&=&32\pi C_F\sqrt{2N_c}M_B^2\int_0^1 [dx]\int_0^{\infty}
b_1 db_1 b_2 db_2\phi_B(x_1,b_1)
\nonumber \\
& &\times \{\left[-r_\pi x_3\phi_K(x_2)\phi'_\pi(x_3)
-r_K(x_1-x_2)\phi'_K(x_2)\phi_\pi(x_3)\right]
E'_{a6}(t^{(1)}_f)h^{(1)}_f(x_1,x_2,x_3,b_1,b_2,b_2)
\nonumber \\
& &-\left[r_\pi(2-x_3)\phi_K(x_2)\phi'_\pi(x_3)
-r_K(2-x_1-x_2)\phi'_K(x_2)\phi_\pi(x_3)\right]
\nonumber\\
& &\hspace{1.0cm}\times
E'_{a6}(t^{(2)}_f)h^{(2)}_f(x_1,x_2,x_3,b_1,b_2,b_2)\}\;,
\label{mf5}
\end{eqnarray}
with the number of colors $N_c=3$, the definition
$[dx]\equiv dx_1dx_2dx_3$, and the evolution factors
\begin{eqnarray}
E'_{ei}(t)&=&\alpha_s(t)a'_i(t)\exp[-S(t)|_{b_3=b_1}]\;,
\\
E'_{ai}(t)&=&\alpha_s(t)a'_i(t)\exp[-S(t)|_{b_3=b_2}]\;.
\end{eqnarray}
The expression of ${\cal M}_e$ (${\cal M}_a$)
is the same as ${\cal M}^P_{e4}$ (${\cal M}^P_{a4}$) but with the Wilson
coefficient $a'_1(t_d)$ ($a'_1(t_f)$). The nonfactorizable amplitude
${\cal M}_{eK}$ is written as
\begin{eqnarray}
{\cal M}_{eK}&=&32\pi C_F\sqrt{2N_c}M_B^2\int_0^1 [dx]\int_0^{\infty}
b_1 db_1 b_3 db_3\phi_B(x_1,b_1)\phi_\pi(x_3)
\nonumber \\
& &\times \{\left[(x_1-x_2-x_3)\phi_K(x_2)
+r_K x_2\phi'_K(x_2)\right]E'_2(t^{(1)}_{dK})
h^{(1)}_d(x_1,x_3,x_2,b_1,b_1,b_3)
\nonumber \\
& &+\left[(1-x_1-x_3)\phi_K(x_2)-r_K x_2\phi'_K(x_2)\right]
E'_2(t^{(2)}_{dK})h^{(2)}_d(x_1,x_3,x_2,b_1,b_1,b_3)\}\;,
\label{mdk}
\end{eqnarray}
with the evolution factor
\begin{eqnarray}
E'_2(t)=\alpha_s(t)a'_2(t)\exp[-S(t)|_{b_2=b_1}]\;.
\end{eqnarray}

The functions $h^{(j)}$, $j=1$ and 2, appearing in
Eqs.~(\ref{md3})-(\ref{mf5}) and in Eq.~(\ref{mdk}), are written as
\begin{eqnarray}
h^{(j)}_d&=& \left[\theta(b_1-b_2)K_0\left(DM_B
b_1\right)I_0\left(DM_Bb_2\right)\right. \nonumber \\
& &\quad \left.
+\theta(b_2-b_1)K_0\left(DM_B b_2\right)
I_0\left(DM_B b_1\right)\right]  
\nonumber \\
&  & \times  K_{0}(D_{j}M_Bb_{2})\;,\;\;\;\;\;\;\;\;\;\;\;\;\;\;  
\mbox{for $D^2_{j} \geq 0$}\;,
\nonumber  \\
&  & \times \frac{i\pi}{2} H_{0}^{(1)}(\sqrt{|D_{j}^2|}M_Bb_{2})\;,\;\;\;\;
 \mbox{for $D^2_{j} \leq 0$}\;,           
\label{hjd}
\\
h^{(j)}_f&=& \frac{i\pi}{2}
\left[\theta(b_1-b_2)H_0^{(1)}\left(FM_B
b_1\right)J_0\left(FM_Bb_2\right)\right. \nonumber \\
& &\quad\left.
+\theta(b_2-b_1)H_0^{(1)}\left(FM_B b_2\right)
J_0\left(FM_B b_1\right)\right]
\nonumber \\
&  & \times
 K_{0}(F_{j}M_Bb_{1})\;,\;\;\;\;\;\;\;\;\;\;\;\;\;\;  
\mbox{for $F^2_{j} \geq 0$}\;,  
\nonumber\\
&  & \times \frac{i\pi}{2} H_{0}^{(1)}(\sqrt{|F_{j}^2|}M_Bb_{1})\;,  
\;\;\;\; \mbox{for $F^2_{j} \leq 0$}\;,           
\label{hjf}
\end{eqnarray}
with the variables
\begin{eqnarray}
D^{2}&=&x_{1}x_{3}\;, 
\nonumber \\
D_{1}^{2}&=&F_1^2=(x_{1}-x_{2})x_{3}\;,
\nonumber \\
D_{2}^{2}&=&-(1-x_{1}-x_{2})x_{3}\;, 
\nonumber \\
F^{2}&=&x_{2}x_{3}\;, 
\nonumber \\
F_{2}^{2}&=&x_{1}+x_{2}+(1-x_{1}-x_{2})x_{3}\;.
\end{eqnarray}
For details of the derivation of $h^{(j)}$, refer to \cite{WYL}.
The hard scales $t^{(j)}$ are chosen as
\begin{eqnarray}
t^{(1)}_d&=&{\rm max}(DM_B,\sqrt{|D_1^2|}M_B,1/b_1,1/b_2)\;,
\nonumber \\
t^{(2)}_d&=&{\rm max}(DM_B,\sqrt{|D_2^2|}M_B,1/b_1,1/b_2)\;,
\nonumber \\
t^{(1)}_f&=&{\rm max}(FM_B,\sqrt{|F_1^2|}M_B,1/b_1,1/b_2)\;,
\nonumber \\
t^{(2)}_f&=&{\rm max}(FM_B,\sqrt{|F_2^2|}M_B,1/b_1,1/b_2)\;,
\nonumber \\
t^{(1)}_{dK}&=&{\rm max}(DM_B,\sqrt{|D_1^2|}M_B,1/b_1,1/b_3)\;,
\nonumber \\
t^{(2)}_{dK}&=&{\rm max}(DM_B,\sqrt{|D_2^2|}M_B,1/b_1,1/b_3)\;.
\end{eqnarray}

In the above expressions the Wilson coefficients are defined by
\begin{eqnarray}
a_1&=&C_2+\frac{C_1}{N_c}\;,
\nonumber\\
a'_1&=&\frac{C_1}{N_c}\;,
\nonumber\\
a_2&=&C_1+\frac{C_2}{N_c}\;,
\nonumber\\
a'_2&=&\frac{C_2}{N_c}\;,
\nonumber\\
a_4&=&C_4+\frac{C_3}{N_c}+\frac{3}{2}e_q
\left(C_{10}+\frac{C_9}{N_c}\right)\;,
\nonumber\\
a'_4&=&\frac{1}{N_c}\left(C_3+\frac{3}{2}e_q C_9\right)\;,
\nonumber\\
a_6&=&C_6+\frac{C_5}{N_c}+\frac{3}{2}e_q
\left(C_{8}+\frac{C_7}{N_c}\right)\;,
\nonumber\\
a'_6&=&\frac{1}{N_c}\left(C_5+\frac{3}{2}e_q C_7\right)\;.
\label{wilc}
\end{eqnarray}
Both QCD and electroweak penguin contributions have been included
as shown in Eq.~(\ref{wilc}). The factors $r_\pi$ and $r_K$,
\begin{eqnarray}
& &r_\pi=\frac{m_{0\pi}}{M_B}\;,\;\;\;\;
m_{0\pi}=\frac{M_\pi^2}{m_u+m_d}\;,
\\
& &r_K=\frac{m_{0K}}{M_B}\;,\;\;\;\;m_{0K}=\frac{M_K^2}{m_s+m_d}\;,
\end{eqnarray}
with $m_u$, $m_d$, and $m_s$ being the masses of the $u$, $d$ and $s$
quarks, respectively, are associated with the normalizations of the
pseudoscalar wave functions $\phi'$.
The pseudovector and pseudoscalar pion wave functions $\phi_\pi$ and
$\phi'_\pi$ are defined by
\begin{eqnarray}
\phi_\pi(x)&=&\int\frac{dy^+}{2\pi}e^{-ixP_3^-y^+}\frac{1}{2}
\langle 0|{\bar d}(y^+)\gamma^-\gamma_5 u(0)|\pi\rangle\;,
\\
\frac{m_{0\pi}}{P_3^-}\phi'_\pi(x)&=&
\int\frac{dy^+}{2\pi}e^{-ixP_3^-y^+}\frac{1}{2}
\langle 0|{\bar d}(y^+)\gamma_5 u(0)|\pi\rangle\;,
\end{eqnarray}
satisfying the normalization
\begin{eqnarray}
\int_0^1 dx\phi_\pi(x)=\int_0^1 dx\phi'_\pi(x)=
\frac{f_\pi}{2\sqrt{2N_c}}\;,
\end{eqnarray}
with the pion decay constant $f_\pi$.
The kaon wave functions $\phi_K$ and $\phi'_K$ possess similar
definitions and normalizations with the $d$ quark field, $m_d$, $M_\pi$
and $f_\pi$ replaced by the $s$ quark field, $m_s$, $M_K$ and $f_K$,
respectively.

Note that we have included the intrinsic $b$ dependence for the heavy
meson wave function $\phi_B$ but not for the light meson wave functions
$\phi_\pi$ and $\phi_K$. It has been shown that the intrinsic $b$
dependence of the light meson wave functions, resulting in only 4\%
reduction of the predictions for the $B\to \pi$ form factor, is not
important \cite{LY1}. It is reasonable to assume that the intrinsic $b$
dependence of the kaon wave function, which is still unknown, is not
essential either. As the transverse extent $b$ approaches zero, the $B$
meson wave function $\phi_B(x,b)$
reduces to the standard parton model $\phi_B(x)$, {\it i.e.},
$\phi_B(x)=\phi_B(x,b=0)$, which satisfies the normalization
\begin{equation}
\int_0^1\phi_B(x)dx=\frac{f_B}{2\sqrt{2N_c}}\;.
\label{dco}
\end{equation}
We do not distinguish the pseudovector and pseudoscalar components
of the $B$ meson wave functions under the heavy quark approximation.
They have roughly the same normalizations because of
$M_B/(m_b+m_d)\sim 1$.

\section{NUMERICAL ANALYSIS}

In the factorization formulas derived in Sec.~IV, the Wilson coefficients
evolve with the hard scale $t$ that depends on the internal kinematic
variables $x_i$ and $b_i$.
Wilson coefficients at a scale $\mu < M_W$ are related to the
corresponding ones at $\mu = M_W$ through usual RG equations.
In our analysis we adopt the leading-order expressions for the
Wilson coefficients with QCD and electroweak penguins included,
\begin{equation}
{\vec C}(\mu) = T_g \left[\exp\left(\int_{g(M_W)}^{g(\mu)} dg^{'}
{\hat{\gamma}^{(0)T}(g^{'}) \over \beta(g^{'})} \right) \right] \cdot
{\vec C}(M_W)\;,
\end{equation}
where the leading-order anomalous dimension matrices
${\hat\gamma}^{(0)}$ are referred to \cite{REVIEW}.
The matching conditions at $\mu = M_W$ \cite{Buchalla} and the choices
of the relevant parameters are given in Appendix A.

Since the typical scale $t$ of a hard amplitude is smaller than the
$b$ quark mass $m_b$, we further evolve the Wilson coefficients from
$\mu=m_b$ down to $\mu = t$ using the RG equation,
\begin{equation}
\mu{d \over d \mu} \vec{C}(\mu) = \left[
{\alpha_s(\mu) \over 4 \pi} {\hat{\gamma}_s^{(0)T}} 
+ {\alpha_{em}(\mu) \over 4 \pi} {\hat{\gamma}_e^{(0)T}} \right]
\cdot \vec{C}(\mu)\;,
\label{RGE-C}
\end{equation}
where the anomalous dimensions ${\hat\gamma}^{(0)}_{s,e}$ for $f = 4$
are referred to \cite{REVIEW}. The solution to Eq.~(\ref{RGE-C})
and the values of the Wilson coefficients $C_i(m_b)$ are also listed in
the Appendix A. For the scale $t$ below the $c$ quark mass $m_c = 1.5$
GeV, we still employ the evolution function with $f = 4$, instead of with
$f =3$, for simplicity, since the matching at $m_c$ is less essential. 
Therefore, we set $f=4$ in the RG evolution between $t$ and $1/b$ 
governed by the quark anomalous dimension $\gamma$.

For the $B$ meson wave function, we adopt the model 
\begin{eqnarray}
\phi_B(x,b)=N_Bx^2(1-x)^2
\exp\left[-\frac{1}{2}\left(\frac{xM_B}{\omega_B}\right)^2
-\frac{\omega_B^2 b^2}{2}\right]\;,
\label{os}
\end{eqnarray}
with the shape parameter $\omega_B=0.4$ GeV \cite{BW}. The 
normalization constant $N_B$, which is related to the decay constant
$f_B$, will be determined below. 
As to the pion wave functions, we employ the models
\begin{eqnarray}
\phi_\pi(x)&=&\frac{3}{\sqrt{2N_c}}f_\pi x(1-x)[1+
c_\pi(5(1-2x)^2-1)]\;,
\label{pwf}\\
\phi'_\pi(x)&=&\frac{3}{\sqrt{2N_c}}f_\pi x(1-x)[1+
c'_\pi(5(1-2x)^2-1)]\;,
\label{wpp}
\end{eqnarray}
with the shape parameters $c_\pi$ and $c'_\pi$.
The kaon wave functions are chosen as
\begin{eqnarray}
\phi_{K}(x)&=&\frac{3}{\sqrt{2N_c}}f_{K}
x(1-x)[1+0.51(1-2x)+0.3(5(1-2x)^2-1)]\;,
\label{mn}\\
\phi'_K(x)&=&\frac{3}{\sqrt{2N_c}}f_{K}
x(1-x)[1+c'_K (5(1-2x)^2-1)]\;.
\end{eqnarray}
$\phi_K$ is derived from QCD sum rules \cite{PB2}, where the second term
$1-2x$, rendering $\phi_K$ a bit asymmetric, corresponds to $SU(3)$
symmetry breaking effect. The decay constant $f_K$ is set to 160 MeV
(in the convention $f_\pi=130$ MeV). Since predictions for the
$B\to K\pi$ decays are insensitive to the kaon wave functions, we
simply adopt the result of QCD sum rules. For the same reason, we 
assume that $\phi'_K$ and $\phi'_\pi$ possess the same functional 
form and that the shape parameter $c'_K$ of the term $5(1-2x)^2-1$ 
in $\phi'_K$ is equal to $c'_\pi$.

We propose to determine $c_\pi$ from the branching ratios of the 
$B\to D\pi$ decays:
\begin{equation}
R_D=\frac{{\rm Br}(B^-\to D^0\pi^-)}{{\rm Br}({\bar B}_d^0\to D^+\pi^-)}\;,
\end{equation}
because this quantity is insensitive to $m_{0\pi}$ and $\phi'_\pi$.
In order to render PQCD predictions reach the central value of the 
data of $R_D=1.61$ \cite{MSA}, a large $c_\pi=0.8$, which enhances
nonfactorizable contributions to the $B^-\to D^0\pi^-$ decay, is
preferred. On the other hand, the data of the $B \to \rho \pi$ decays
also imply a large $c_\pi$. To further enhance nonfactorizable
contributions relative to factorizable ones, the $B$ meson wave function
with $\phi_B \to x^2$ as $x\to 0$ has been assumed as shown in
Eq.~(\ref{os}). This behavior, different from that of the model
$\phi_B \to \sqrt{x}$ as $x\to 0$ proposed in \cite{BW}, decreases
factorizable contributions. Note that nonfactorizable contributions are
insensitive to the variation of the $B$ meson wave funciton.
The details for the above numerical study will be published elsewhere. 
The extracted pion wave function $\phi_\pi$ with $c_\pi=0.8$
is close to the Chernyak-Zhitnitsky model with $c_\pi=1.0$ \cite{CZ}.
It differs from the asymptotic model with $c_\pi=0$, which has been
extracted from the data of the pion transition form factor involved in 
the process $\pi\gamma^*\to\gamma$ \cite{KM}. We shall argue that
$\pi\gamma^*\to\gamma$ is a special process, whose infrared divergences
differ from those of processes containing final-state hadrons \cite{L10}.
Hence, there is no contradiction between $\phi_\pi$ determined from the
$B\to D\pi$ decays and from the pion transition form factor.

We then extract $c'_\pi$ from the data of the pion form factor,
whose factorization formula is written as \cite{Ukai}
\begin{eqnarray}
F_\pi(Q^2)&=&16\pi C_FQ^2\int_0^1 dx_1 dx_2 \int_0^\infty
b_1 db_1 b_2 db_2 \alpha_s(t)\exp[-S_{\pi\pi}(t)]
\nonumber \\
& &\times x_2\left[\phi_\pi(x_1)\phi_\pi(x_2)
+2r_\pi\phi'(x_1)\phi'_\pi(x_2)
\right]h(x_1,x_2,b_1,b_2)\;,
\end{eqnarray}
with
\begin{eqnarray}
S_{\pi\pi}(t)&=&s(x_1P_{\pi 1}^+,b_1)+s((1-x_1)P_{\pi 1}^+,b_1)+
s(x_2P_{\pi 2}^-,b_2)+s((1-x_2)P_{\pi 2}^-,b_2)
\nonumber\\
& &+2\int_{1/b_1}^{t}
\frac{d{\bar \mu}}{\bar \mu}\gamma(\alpha_s({\bar \mu}))
+2\int_{1/b_2}^{t}
\frac{d{\bar \mu}}{\bar \mu}\gamma(\alpha_s({\bar \mu}))\;,
\label{spp}\\
h(x_1,x_2,b_1,b_2)&=&K_{0}\left(\sqrt{x_1x_2}Qb_1\right)
\nonumber \\
& &\times \left[\theta(b_1-b_2)K_0\left(\sqrt{x_2}Q
b_1\right)I_0\left(\sqrt{x_2}Qb_2\right)\right.
\nonumber\\
& &\left.+\theta(b_2-b_1)K_0\left(\sqrt{x_2}Qb_2\right)
I_0\left(\sqrt{x_2}Qb_1\right)\right],
\label{dhpi}\\
t&=&{\rm max}(\sqrt{x_1x_2}Q,1/b_1,1/b_2)\;.
\end{eqnarray}
The momentum transfer is defined by $Q^2=2P_{\pi 1}\cdot P_{\pi 2}$,
$P_{\pi 1}$ and $P_{\pi 2}$ being the momenta of the intial and final
pions, respectively. Useful references for the derivation
of the above expression are \cite{LS,TL,CDH}. The data are
$Q^2F_\pi(Q^2)\sim 0.4\pm 0.2$ GeV$^2$ for $Q^2 > 4$ GeV$^2$
\cite{JB,SRA}. Adopting the quark masses $m_u=4.5$ MeV and
$m_d=1.8 m_d=8.1$ MeV, which lead to $m_{0\pi}=1.53$ GeV, we find that
the choice $c'_\pi=0$ gives the pion form factor
$Q^2F_\pi(Q^2)\sim 0.4$ GeV$^2$ for $Q^2=6.3$ GeV$^2$. Hence, we
choose $c'_K=c'_\pi=0$ as stated before.

With the pion wave functions fixed in the above procedures, we determine
the $B$ meson decay constant $f_B$ (or $N_B$ in Eq.~(\ref{os})) from the
data of the $B_d^0\to\pi^\pm\pi^\mp$ decay \cite{CLEO3},
\begin{equation}
{\rm Br}(B_d^0\to \pi^\pm\pi^\mp)=
(4.3^{+1.6}_{-1.4}\pm 0.5) \times 10^{-6}\;,
\label{dpp}
\end{equation}
where ${\rm Br}(B_d^0\to \pi^\pm\pi^\mp)$ represents the CP average of
${\rm Br}(B_d^0\to \pi^+\pi^-)$ and ${\rm Br}({\bar B}_d^0\to \pi^-\pi^+)$.
We employ $G_F=1.16639\times 10^{-5}$ GeV$^{-2}$, the Wolfstein
parameters $\lambda=0.2196$, $A=0.819$, and $R_b=0.38$, the masses 
$M_B=5.28$ GeV, and the ${\bar B}_d^0$ ($B^-$) meson lifetime 
$\tau_{B^0}=1.55$ ps ($\tau_{B^-}=1.65$ ps) \cite{LEP}. For the 
factorization formulas of the $B\to \pi\pi$ decays, refer to \cite{LUY}. 
For the angle $\phi_3= 90^o$, the result is $f_B=190$ MeV, which
corresponds to ${\rm Br}(B_d^0\to \pi^\pm\pi^\mp)=4.5 \times 10^{-6}$
and the $B\to\pi$ transition form factor
\begin{equation}
F^{B\pi}(q^2=0)=0.25\;,
\end{equation}
which is also reasonable. Here $q$ stands for the momentum carried away
by the external $W$-emission. The value of $f_B$ is close to that adopted
in the PQCD studies of the $B\to D\pi$ and $B\to K^*\gamma$ decays
\cite{LL,LM}, and consistent with those from lattice calculations 
\cite{AK} and from QCD sum rules \cite{SN} in the literature. The 
motivation to choose $\phi_3=90^o$ will be
explained later.

We emphasize that the decay constant $f_B$ can not be determined
unambigiously in the current analysis. The above value $f_B=190$ MeV
corresponds to the shape parameter $\omega_B=0.4$ GeV. Changing
$\omega_B$, different $f_B$ will be obtained when fitting PQCD
predictions to the data in Eq.~(\ref{dpp}). However, if more data, such
as the CP asymmetry in the $B_d^0\to \pi^\pm\pi^\mp$ decays, are
available, both $\omega_B$ and $f_B$ can be uniquely determined. The
reason is that tree and penguin contributions depend on $\omega_B$ and
$f_B$ simultaneously, while annihilation contributions, the most
important source of strong phases as shown below, depend only on $f_B$.
Because the branching ratio, mainly determined by tree and penguin
contributions, and the CP asymmetry, related to strong phases of
annihilation contributions, vary with the $B$ meson wave function in a 
different way, their data can fix $\omega_B$ and $f_B$ uniquely.

Note that the above parameters are obtained by fitting
predictions to the central values of the available data. If taking into
account the uncertainty of the data, the allowed range of the parameters
is in fact huge. For example, any value of the shape parameter
$c_\pi$ in the pion wave funciton $\phi_\pi$ between 0.4 and 1.0
is acceptable for the data of $R_D$. The shape parameter $c'_K$ in
the pseudoscalar kaon wave function $\phi'_K$ can differ from $c'_\pi$
in $\phi'_\pi$. In this work we do not intend to determine the range
of parameters, but adopt representative parameters to make predictions
for the $B\to K\pi$ decays, and examine whether the predictions are
consistent with the data. For a summary of the parameters we have
adopted in the numerical analysis, refer to Appendix B.

With all the meson wave functions fixed, we predict the branching
ratios and the CP asymmetries of the $B\to K\pi$ decays. The $s$ quark
mass is set to $m_s=100$ MeV, which corresponds to $m_{0K}=2.22$ GeV. We
derive the branching ratios of the four $B\to K\pi$ modes in
Eq.~(\ref{bkpi}) for different $\phi_3$, which are shown in Fig.~5.
The branching ratios of the $K^{\mp}\pi^0$ and $K^{\mp}\pi^{\pm}$
modes increase with $\phi_3$, while those of the $K^0\pi^{\pm}$ and
$K^0\pi^0$ modes are insensitive to the variation of $\phi_3$. The
increase with $\phi_3$ is mainly a consequence of the inteference between
the penguin contribution $F_e^P$ and the tree contribution $F_e$.
Predictions for the ratio $R$ in Eq.~(\ref{tr}) and the CP asymmetries
$A_{CP}$ in Eqs.~(\ref{cp1})-(\ref{cp}) for different $\phi_3$ are
displayed in Fig.~6 and Fig.~7, respectively. The prediction of $R$
increases from 0.7 to 1.2 when $\phi_3$ moves from 0 to $180^o$.
Unfortunately, the large uncertainty of the current data does not give a
constraint of $\phi_3$. 
Comparing with the central value of the CLEO data of $R$ in
Eq.~(\ref{cld}), we extract $\phi_3= 90^o$. The data of $A_{CP}$ have
also large uncertainties, and do not constraint $\phi_3$ either. Our
analysis shows that the magnitude of $A_{CP}^c$ and $A_{CP}^{'0}$ 
is negligible, smaller than 3\%, while the magnitude of $A_{CP}^0$ and
$A_{CP}^{'c}$ can reach 20\%.

Our predictions for the branching ratio of each mode
corresponding to $\phi_3=90^o$,
\begin{eqnarray}
& &{\rm Br}(B^+\to K^0\pi^+)=20.22\times 10^{-6}\;,
\nonumber\\
& &{\rm Br}(B^-\to \bar{K}^0\pi^-)=19.79\times 10^{-6}\;,
\nonumber\\
& &{\rm Br}(B_d^0\to K^+\pi^-)=22.74\times 10^{-6} \;,
\nonumber\\
& &{\rm Br}({\bar B}_d^0\to K^-\pi^+)=15.50\times 10^{-6}\;,
\nonumber\\
& & {\rm Br}(B^+\to K^+\pi^0)= 11.40\times 10^{-6}\;,
\nonumber\\
& &{\rm Br}(B^-\to K^-\pi^0)=7.89\times 10^{-6}\;,
\nonumber\\
& &{\rm Br}(B_d^0\to K^0\pi^0)= 8.81\times 10^{-6} \;,
\nonumber\\
& &{\rm Br}({\bar B}_d^0\to \bar{K}^0\pi^0)= 9.25\times 10^{-6}\;,
\label{pqp}
\end{eqnarray}
are consistent with the CLEO data \cite{YK},
\begin{eqnarray}
& &{\rm Br}(B^\pm\to K^0\pi^\pm)
=(18.2^{+4.6}_{-4.0}\pm 1.6)\times 10^{-6}\;,
\nonumber\\
& &{\rm Br}(B_d^0\to K^\pm\pi^\mp)
=(17.2^{+2.5}_{-2.4}\pm 1.2)\times 10^{-6} \;,
\nonumber\\
& &{\rm Br}(B^\pm\to K^\pm\pi^0)=(11.6^{+3.0+1.4}_{-2.7-1.3})
\times 10^{-6}\;,
\nonumber\\
& &{\rm Br}(B_d^0\to K^0\pi^0)=(14.6^{+5.9+2.4}_{-5.1-3.3})
\times 10^{-6} \;.
\end{eqnarray}
To derive Eq.~(\ref{pqp}), we have distinguished the pion masses
$M_{\pi^\pm}=139$ MeV and $M_{\pi^0}=135$ MeV.

The PQCD results of each form factor and nonfactorizable amplitude
involved in the $B^{0} \to K^+\pi^-$ decay are listed in Table I. It
indicates that nonfactorizable contributions are only few percents of
factorizable ones. This is the reason FA works well for most
two-body $B$ meson decay modes. However, there are exceptions. For
modes whose factorizable contributions are proportional to the
small Wilson coefficient $a_2$, such as $B\to J/\psi K^{(*)}$,
nonfactorizable contributions become important. Similarly, the term
$F_{eK}$, proportional to $a_2$, is small. Hence, the branching ratios
of the $K\pi^0$ modes are about half of those of the $K\pi^\pm$ modes.
Table I also indicates that
the factorizable annhilation diagrams contribute 
dominant strong phases. The reason has been discussed in
\cite{KLS}. If expressing the amplitude of the $B_d^{0} \to K^+\pi^-$
decay as
\begin{equation}
{\cal A} \simeq V_{t}^{*} P e^{i\delta_{P}} -V_{u}^{*} T,
\end{equation}
with the penguin contribution $P = |f_K F_e^{P} + f_B F_a^{P}|$
and the tree contribution $T = |f_K F_e|$,
the strong phase $\delta_P$ is as large as
\begin{equation}
\delta_{P} = 144^o\;.
\end{equation}
This result is consistent with the conclusion drawn from a global fit
to data of two-body charmless $B$ meson decays \cite{HY2}, where the
strong phase was introduced as a free parameter.

To test the sensitivity of our predictions to different choices of
model wave functions and parameters, we have varied the shape parameter
$\omega_B$ for the $B$ meson wave funciton from 0.3 to 0.5,
the shape parameter $c'_K$ for the kaon wave funciton from
0 to 0.8, the masses $m_{0K(\pi)}$ from 1.3 GeV to 2.7 GeV,
the forms of the meson wave functions, such as
\begin{eqnarray}
\phi_B^{\rm test}(x,b)&=&N_B^{\rm test}x(1-x)
\exp\left[-\frac{1}{2}\left(\frac{xM_B}{\omega_B}\right)^2
-\frac{\omega_B^2 b^2}{2}\right]\;,
\\
\phi_{K}^{\rm test}(x)&=&\frac{\sqrt{6}}{2}f_{K}
x(1-x)[1+0.3(5(1-2x)^2-1)]\;,
\end{eqnarray}
for the $B$ meson and the kaon, and the asymptotic model 
\begin{eqnarray}
\phi_\pi^{\rm AS}(x)=\frac{3}{\sqrt{2N_c}}f_\pi x(1-x)\;,
\end{eqnarray}
for the pion, and the Wolfstein parameters $\lambda$ from
0.21 to 0.22. It is found that our predictions for $R$
change by less than 5\%, and are very stable. That is, $R$ is an 
appropriate quantity for the determination of $\phi_3$,

There are other theoretical uncertainties from
higher-order $O(\alpha_s^2)$ and higher-twist $O(1/M_B)$ corrections.
For a simple estimation, we examine the fractional contribution to
the form factor $F^{B\pi}$ as a function of 
$\alpha_s(t)/\pi$. It is observed that 90\% and 97\% of the 
contributions arise from the region with $\alpha_s(t)/\pi<0.2$ 
and with $\alpha_s(t)/\pi<0.3$, respectively. Therefore, our PQCD 
results are well within the perturbative region. It is reasonable
to assume that $O(\alpha_s^2)$ corrections to the decay amplitudes
are below 20\%. In the derivation of the hard functions, we have 
neglected the mass difference ${\bar\Lambda}=M_B-m_b$ to obtain the
leading-twist factorization formulas. Next-to-leading-twist
corrections, proportional to ${\bar\Lambda}/M_B$, are then about 10\%.

\section{PENGUIN ENHANCEMENT}

In this section we shall highlight the enhancement of penguin
contributions observed in the PQCD approach, and its role in the
explanation of the $B\to \pi\pi$ and $B\to K\pi$ data. For simplicity, we
demonstrate our observation by means of the FA approach.
Consider the ratios $R$ in Eq.~(\ref{tr}) and $R_\pi$ defined by
\begin{equation}
R_\pi=\frac{{\rm Br}(B_d^0 \to  K^\pm \pi^\mp)}
{{\rm Br}(B_d^0 \to \pi^\pm \pi^\mp)}\;,
\end{equation}
which can be written as
\begin{eqnarray}
R&=& \frac{ a_K^2 + 2 a_K \lambda^2 R_b \cos\phi_3}{a_K^2}\;,
\\
R_\pi&= &\frac{ a_K^2 + 2 a_K \lambda^2 R_b \cos\phi_3}
           {\lambda^2 R_b [R_b +2 a_\pi (R_b -\cos\phi_3)]}\;.
\label{rkpi}
\end{eqnarray} 
The factors
\begin{eqnarray}
a_K=\frac{a_4 + 2 r_K a_6}{a_1}\;,\;\;\;\;   
a_\pi=\frac{a_4 + 2 r_\pi a_6}{a_1}\;,
\end{eqnarray}
being negative values, represent the ratios of the penguin contribution
to the tree contribution in the $K\pi$ and $\pi\pi$ modes, respectively.
It is obvious that the data $R\sim 1$ imply $\phi_3 \sim 90^o$, no matter
what $a_K$, $\lambda$ and $R_b$ are. It is the reason when we vary all
the parameters in the analysis in Sec.~IV, the extraction of $\phi_3$
remains invariant.

While to determine $\phi_3$ from the data of the ratio $R_\pi\sim 4$, one
must have precise information of $a_K$ and $a_\pi$, and of the parameters
$\lambda$ and $R_b$. It can be shown that the extraction of $\phi_3$
from $R_\pi$ depends on these parameters sensitively. Hence, $R_\pi$ is
not an appropriate quantity for the determination of $\phi_3$. To explain
the data of $R_\pi\sim 4$ in the FA approach, an unreasonably large 
$m_0\sim 4$ GeV corresponding to $m_d=2m_u=3$ MeV, {\it i.e.}, large
$|a_{K(\pi)}|\sim 0.09$ and a large $\phi_3\sim 130^o$ must be postulated
\cite{WS}. This is obvious from Eq.~(\ref{rkpi}), since a large
$|a_{K(\pi)}|$ enhances $R_\pi$, and a large $\phi_3$ leads to
constructive interference between the two terms
in the numerator of $R_\pi$. The determination $\phi_3\sim 114^o$ from
global fits to charmless $B$ meson decays \cite{WS}, located between the
two extreme cases $90^o$ and $130^o$, is then understood. The result of
$\phi_3$ will become even larger, if reasonable quark masses
$m_d=2m_u\sim 10$ MeV are employed. The huge difference between
$90^o$ and $130^o$ extracted from different data renders the 
determination of $\phi_3$ in the
FA approach less convincing. In the modified FA approach with
effective number of colors $N_c^{\rm eff}$, a large unitarity angle
$\phi_3\sim 105^o$ is also concluded \cite{HYCheng}.

An interesting question is as follows. If we give higher weight
to the extraction of $\phi_3$ from $R$, which is more 
model-independent than that from $R_\pi$, can we explain the data 
$R_\pi\sim 4$ using a smaller $\phi_3$? The answer is positive in
the PQCD approach. Table I shows that the ratio of the penguin 
contribution to the tree contribution reaches
\begin{equation}
|a_K|=\left|\frac{F_e^P}{F_e}\right|\sim 0.1\;,
\label{ptt}
\end{equation}
even with a reasonable value of $m_0= 1.53$ GeV. The reason is that we
do not assume the same form factors for the operators $O_{1,2,3,4}$ and
for $O_{5,6}$. These form factors, evaluated explicitly in the PQCD
formalism, possess different factorization formulas as shown in
Eqs.~(\ref{int4}) and (\ref{int6}). It is easy to observe that the
integrands in the two factorization formulas become identical, if the
terms associated with the pseudoscalar wave function $\phi'_\pi$ and the
factors $x_3$ are dropped. The $x_3\to 0$ limit corresponds to the
kinematic configuration, in which the light quark emitted from the $b$
quark decay vertex carries the full meson momentum. This is the
configuration, on which the equality of the two form factors in the FA
approach is constructed. Therefore, the larger ratio of the penguin
contribution to the tree contribution is achieved dynamically, instead
of by increasing $m_0$. With this penguin enhancement, the observed
branching ratios of the $B\to K\pi$ and $B\to\pi\pi$ decays and
$R_\pi\sim 4$ can be explained simultaneously in the PQCD approach
using $m_0= 1.53$ GeV and a smaller $\phi_3=90^o$. That is, the data of
$R_\pi$ do not demand large $m_0$ and $\phi_3$. Such a dynamical
enhancement of penguin contributions can not be obtained in the FA
appraoch.

One of the sources responsible for the penguin enhancement is the RG
evolution effect caused by the running hard scale $t$. In Fig.~8 we
display the RG evolution of the Wilson coefficients $a_{i}(\mu)$,
$i=1,4,6$. It is found that $a_1$ is almost constant for $\mu=500$ MeV to
$M_B$. In contrast, $|a_4|$ and $|a_6|$ dramatically increase as $\mu$
evolves to below $M_B/2$. If choosing $t=M_B/2$ with $m_0=1.5$ GeV, the
ratio $|a_{K(\pi)}|\sim 0.06$, close to that in the FA approach with the
same value of $m_0$, is too small to explain $R_\pi\sim 4$. As stated
before, PQCD provides a prescription for choices of the hard scales $t$:
$t$ should be chosen as the virtualities of internal particles in
Eq.~(\ref{et}) in order to decrease higher-order corrections. It reflects
the fact that energy releases and evolution effects involved in different
$B$ meson decay modes are different. These hard scales can then reach
lower values, at which $|a_6(t)|$ is enhanced over $|a_6(M_B/2)|$. This
evolution effect increases $|a_{K(\pi)}|$ by about 50\% as indicated by
Eq.~(\ref{ptt}).

The enhancement due to the increase of $C_6(t)$ with decreasing
$t$ makes us worry that the contribution from the small $t$ region
may be important. This will invalidate the perturbative expansion of
the hard amplitudes. As a check, we examine the fractional
contribution to $F_{e6}^P$ as a function of $\alpha_s(t)/\pi$. The
results are displayed in Fig.~9, which indicate that about 80\% (90\%) 
of the contributions come from the region with 
${\alpha_s(t)}/{\pi}<0.2$ (0.3). Therefore, exchanged gluons are still 
hard enough to guarantee the applicability of PQCD.

Another source of penguin enhancement is the behavior of the $B$ meson
wave function at $x\to 0$. As shown in Eqs.~(\ref{int4}) and (\ref{int6}),
the factorization formulas consist of two terms. It can be easily
verified that when the two terms are roughly equal, the ratio of the
penguin contribution to the tree contribution reaches its maximum. A
simple investigation reveals the approximate expressions of the hard
functions at small momentum fractions,
\begin{eqnarray}
& &h_e(x_1,x_3,b_1,b_3)\sim \ln(x_1x_3)\ln x_3\;,
\nonumber\\
& &h_e(x_3,x_1,b_3,b_1)\sim \ln(x_1x_3)\ln x_1\;.
\end{eqnarray}
A $B$ meson wave function with other behaviors, say, $\phi_B\sim x$ or
$\sqrt{x}$ \cite{BW} as $x\to 0$, leads to the dominance of the
second term, and the penguin contribution becomes relatively smaller.
While the $B$ meson wave function in Eq.~(\ref{os}), which vanishes like
$x^2$ as $x\to 0$, renders the contributions from the above two terms
approximately the same. The penguin contribution corresponding to
Eq.~(\ref{os}) is about 10\% larger than that corresponding to the
model in \cite{BW}.

\section{FINAL STATE INTERACTION}

Two-body FSI effects have been studied in various ways \cite{FSI}. It
was found that these effects enhance the CP asymmetry in the
$B^{\pm} \to K^{0}\pi^{\pm}$ modes from order 0.5 \% under the
FA \cite{KPS} up to order (10 - 20) \%. However,
Kamal has pointed out that the large CP asymmetry is due to an
overestimation of FSI effects by a factor of 20 \cite{Kamal}. For a
critic assessment on the analyses of FSI effects in the literature, refer
to \cite{Kamal}. 

We briefly sketch the methods used in most of the estimates of FSI
effects. For simplicity, we consider only the $B^+\to K^0\pi^+$ decay.
The unitarity relation for the amplitude ${\cal A}(B^+\to K^0\pi^+)$ is
written as
\begin{equation}
\Im {\cal A}(B^+\to K^0\pi^+)=\frac{1}{2}
\sum_N2\pi\delta(M_B-E_N) {\cal A}(N\to K^0\pi^+){\cal A}^*(B^+ \to N)\;.
\label{ur}
\end{equation}
If only the elastic channel $K^+\pi^0$ contributes, Watson's theorem
tells that the phase of ${\cal A}(B^+\to K^0\pi^+)$ is given by the $S$
wave $I=3/2$ phase shift. This argument works for the $K\to\pi\pi$ decays
but not for $B$ meson decays. For $M_B\sim 5$ GeV, many channels
contribute and Watson's theorem says nothing about the strong phase of
${\cal A}(B^+\to K^0\pi^+)$. In fact, even if the phases of
${\cal A}(N\to K^0\pi^+)$ for all $N$ are known, the unitarity relation
does still not fix the phase of $A(B^+ \to K^0\pi^+)$ uniquely.

Inspite of this difficulty, some authors computed
${\cal A}(N\to K^0\pi^+)$ for few $N$. Certainly, more than the unitarity
relation is needed to obatin the strong phase of $A(B^+\to K^0\pi^+)$.
The phases of ${\cal A}(N\to K^0\pi^+)$ are often estimated by
a Regge analysis. However, this method is reliable only near the forward 
direction. In our problem we need $S$ wave amplitudes, {\it i.e.},
scattering amplitudes for all angles. A big assumption of a
straight line trajectory has been adopted. This is highly questionable,
especially for Pomerons. For these reasons, we believe that the above
analyses are qualitative at most.

It is our viewpoint that if a strong phase can not be determined in QCD,
there is no other way to compute it. A simple physical picture of FSI,
the color-transparency argument, has been put forward by Bjorken
\cite{Bjorken}:
\begin{quotation}
Since products of a $B$ meson decay into two light mesons are quite
energetic, the quark-antiquark pair inside a meson remains a state of
small size with a correspondingly small chromomagnetic moment until it is
far from the other meson. It is then more realistic that the two quark
pairs group individually into final-state mesons without further
exchanging soft gluons. 
\end{quotation}
This picture is consistent with our observation:
Sudakov suppression is strong for large meson momenta as shown in
Eq.~(\ref{sbk}), which then demands final-state mesons of small
transverse extent $b$, {\it i.e.}, of small chromomagnetic moment.

The effects from soft gluon exchanges among mesons in two-body
heavy meson decays have been analyzed quantitatively by means of RG
methods, which sum up large logarithms produced by infinite gluon
emissions. It was found that these effects generate only small FSI
phases for $B$ meson decays, in agreement with the color-transparence
argument, but large FSI phases for $D$ meson decays \cite{LT}. 
That soft gluon effects are large in $D$ meson
decays is expected, since Sudakov suppression is weaker, two quarks in a
final-state meson is separated by a larger distance, and soft gluons can
resolve the color structure of that meson. Based on the above 
reasonings, we have neglected FSI effects in the PQCD approach
to two-body $B$ meson decays.

To justify the neglect of FSI, we apply our formalism to the 
$B\to\pi\pi$ and $B\to K\pi$ decays without taking into account these
long-distance effects. FSI in these decays should be different. 
Since tree contributions dominate in the $B\to\pi\pi$ decays, extra 
phases from FSI do not change branching ratios very much. This argument 
applies to the decays $B^\pm \to K^0\pi^\pm$ and $B_d^0\to K^0\pi^0$, 
where penguin contributions dominate. While the 
inteference between tree and penguin contributions plays an essential 
role in the $B_d^0\to K^\pm\pi^\mp$ and $B^\pm \to K^\pm \pi^0$ decays.
Large FSI effects will change the 
relative phases between tree and penguin amplitudes, and thus branching 
ratios. If the same formalism without including FSI can be applied to 
both decays successfully, we believe that these long-distance effects 
are negligible. The agreement of our results with the data shown in 
Eq.~(\ref{pqp}) implies this conclusion.

Without FSI, large strong phases come from the factorizable annihilation
diagrams (phases from nonfactorizable diagrams are small) in the PQCD
approach as stated in Sec.~IV.
There has been a widely spread folklore that the annihilation diagrams
give negligible contribution due to helicity suppression, 
the same as in $\pi\to e\overline\nu$ decay. That is, 
a left-handed massless electron and a right-handed antineutrino
can not fly away back to back because of angular momentum conservation.
However, this argument does not apply to $F_{a6}^P$. A left-handed quark
and a left-handed antiquark, for which helicities are dictated by the
$O_6$ operator, can indeed fly away back to back \cite{chang}.
These behaviors have been reflected by Eqs.~(\ref{exc4}) and (\ref{exc6}):
Eq.~(\ref{exc4}) vanishes exactly, if the kaon and pion wave functions
are identical, while the two terms in Eq.~(\ref{exc6}) are constructive.
The reason the annihilation diagrams from the $O_6$ operator possess
large absorptive parts can be understood in the following way. The cuts
on the internal quark lines in Figs.~3(e) and 3(f) correspond to a
process
\begin{equation}
B^+\to {\bar s} u\to K^0\pi^+\;.
\end{equation}
The intermediate state $(\sbar u)$ can be regarded being highly inelastic,
if expanded in terms of hadron states. According to Eq.~(\ref{ur}),
large strong phases are expected.


\section{COMPARISION WITH OTHER ANALYSES}

Recently, Beneke {\it et al.} proposed to evaluate nonfactorizable
contributions to charmless $B$ meson decays in the PQCD framework
\cite{BBNS}. They argued that factorizable contributions (transition
form factors) are not calculable in perturbation theory, but
nonfactorizable contributions are. The reasoning is as follows. The
internal $b$ quark in the hard amplitude may go onto mass shell,
producing a power divergent factor $x^{-2}$, $x$ being the momentum
fraction associated with the pion. The soft divergence from $x\to 0$
can not be removed by a pion wave funciton, if it vanishes like
$x$ as $x\to 0$. Since this divergence is not of the pinched type
which is absorbed into a wave function, its appearence implies the
breakdown of PQCD factorization theorem. While such a power divergence
does not exist in nonfactorizable amplitudes \cite{BBNS}. We argued that
the $x^{-2}$ factor in fact can be easily smeared out by parton
transverse momenta $k_T$ considered in this work or killed by a wave
function vanishing faster than $x$ as $x\to 0$. With the inclusion of
$k_T$ and Sudakov suppression, we have explicitly shown that almost
100\% of the full contribution to the $B\to\pi$ transition form factor
arises from the region with the coupling constant $\alpha_s/\pi <0.3$. 
It indicates that dynamics from hard gluon exchanges indeed
dominate in the PQCD calculation.

There are other important differences between our approach and
\cite{BBNS}. The momentum of the light spectator quark in the $B$ meson
has been ignored in the formalism of \cite{BBNS}, such that quark
propagators in hard nonfactorizable decay amplitudes always remain
time-like. The annihilation diagrams were not included either. With
these approximations, leading-order information of strong phases was lost.
Strong phases then arise from diagrams of the BSS mechanism, which,
as shown below, are small compared to those from 
annihilation diagrams. On the other hand, Sudakov resummation of
large logarithmic corrections was not taken into account. It is then 
expected that higher-order corrections will be large and spoil the 
perturbative expansion. It has been shown \cite{IL} that the PQCD 
formalism without including Sudakov suppression is not applicable to 
exclusive processes for energy scale below 10 GeV.

We show that strong phases from the BSS
mechanism are suppressed by the charm mass threshold and by
$O(\alpha_s)$, since there must be a hard gluon emitted by the spectator
as shown in Fig.~10, which turns the soft spectator in the $B$ meson into
a fast spectator in the final-state meson. That is, the contributions
from the penguin diagram have been overestimated. The charm quark loop
contributes an imaginary part,
\begin{equation}
C_2(t)\alpha_s(t)\int du u(1-u)\theta(q^2 u(1-u)-m_c^2)\;,
\label{stp}
\end{equation}
where $q^2$ is the invariant mass of the gluon emitted from the penguin.
The contribution from the $u$ quark loop is suppressed by the
small CKM factor $|V_u|$. Since $q^2$ is not clearly defined in
the FA approach, it is usually chosen as $q^2=m_b^2$ or
$q^2=m_b^2/2$, and Eq.~(\ref{stp}) gives a substantial amount of
imaginary contribution to decay amplitudes \cite{HY}.

However, the invariant mass $q^2$ can be defined unambigiously in the
PQCD formalism by 
\begin{equation}
q^2=(x_2 P_2+x_3P_3)^2=x_2 x_3 M_B^2\;,
\label{q2p}
\end{equation}
since the quark going into the kaon (pion) carries the fractional
momentum $x_2P_2$ ($x_3P_3$). Then, $q^2=m_b^2$ or $m_b^2/2$
corresponds to a configuration, in which the two quarks produced from
the gluon carry the full momenta of the two final-state mesons.
Obviously, this configuration is unlikely because of the strong
suppression from the kaon and pion wave functions in the large $x$
region. Substituting Eq.~(\ref{q2p}) into Eq.~(\ref{stp}),
an exact numerical analysis indicates that the BSS mechanism contributes
an imaginary part smaller than that from the nonfactorizable
and annihilation amplitudes by a factor of 10. Table II shows how
the imaginay part of the charm quark loop contribution vanishes
with the decarease of $q^2$.

On the issue of FSI, Suzuki has argued that strong phases of the
$B\to K\pi$ amplitudes can not be evaluated in QCD \cite{suzuki}. He
pointed out that the invariant masses of the ${\bar s} u$ and
${\bar d} u$ pairs in Figs.~3(c) and 3(d), respectively, are of order
$(\Lambda_{\rm QCD}M_B)^{1/2}\sim 1.2$ GeV. It implies that the
$B\to K\pi$ decays are located in the resonance region and their strong
phases are very complicated. We have computed the average hard scales
of the $B\to K\pi$ decays, which are about 1.4 GeV, in agreement with 
the above estimate. However, the outging quark pairs possess an invariant 
mass larger than 1.4 GeV, such that the processes are in fact not so 
close to the resonance region. We could interpret that the decays occur
via a six-fermion operator within space smaller than $(1/1.4)$ GeV$^{-1}$.
While they are not completely short-distance, the fact that over 90\% of
contributions come from the $x$-$b$ phase space with 
$\alpha_s(t)/\pi< 0.3$ allows us to estimate the decay amplitudes 
reliably. We believe that the strong phases can be computed up to about 
20\% uncertainties, which result in 30\% errors in the predictions for 
CP asymmetries.

\section{CONCLUSION}

In this paper we have analyzed the $B\to K\pi$ decays using PQCD
factorization theorem. In this approach hadronic matrix elements,
including factorizable and nonfactorizable, and real and imaginary
contributions can be evaluated explicitly. The strong phases arise from
non-pinched singularities of quark and gluon propagators in annihilation
and nonfactorizable hard amplitudes. It has been explicitly shown that
strong phases from the BSS mechanism are small. The analysis of soft
gluon effects and the simultaneous success of the PQCD applications to
the $B\to K\pi$ and $B\to \pi\pi$ decays implied that long-distance
FSI effects are negligible. The universal meson wave functions have been
determined from the available data of the pion form factor and of the
$B\to D\pi$ and $B\to\pi\pi$ decays.
The dependences of the ratio $R$ of the neutral $B$
decay branching ratio to the charged $B$ decay branching ratio and of
the CP asymmetries on $\phi_3$ have been derived.
Our predictions for all the
$B\to K\pi$ modes are consistent with the experimental data.

In spite of potential theoretical uncertainties, we have extracted
the following features for the $B\to K\pi$, $\pi\pi$ decays, which are
less ambiguous:
\begin{enumerate}
\item
Nonfactorizable amplitudes are negligible.
\item
Annihilation diagrams are not negligible.
\item
Annihilation diagrams generate large strong phases.
\item
More precise data are needed in order to obtain a strong
constraint on $\phi_3$.
\item
$R$ is an ideal quantity for the determination of $\phi_3$, since it
is insensitive to all the Wolfstein and nonperturbative QCD parameters.
\item
$\phi_3$ is about $90^o$ from fitting our predictions to the central
value of the data of $R$.
\item
Penguin amplitudes are dynamically enhanced, and larger than
those employed in the FA approach by 50\%.
\item
The data of $B\to\pi\pi$ decays, {\it i.e.}, the ratio
$R_\pi$ of the $B\to K\pi$ branching ratio to the $B\to\pi\pi$ branching
ratio can be explained by the smaller angle $\phi_3\sim 90^o$.
That is, the data of $R_\pi$ do not demand a large $\phi_3 > 90^o$.
\end{enumerate}

\vskip 1.0cm

\centerline{\bf Acknowledgement}
\vskip 0.5cm

We acknowledge useful discussions with members of our PQCD group: E. Kou,
T. Kurimoto, C.-D. L\"u, T. Morozumi, N. Sinha, R. Sinha, K. Ukai, M.Z. Yang
and T. Yoshikawa. This work was supported
in part by Grant-in Aid for Special Project Research (Physics of CP
Violation) and by Grant-in Aid for Scientific Exchange from the Ministry
of Education, Science and Culture of Japan.
The work of HNL was supported by the National Science Council of
R.O.C. under the Grant No. NSC-89-2112-M-006-004. 
The work of YYK was in part supported by 
the National Science Council of R.O.C. under the
Grant No. NSC-89-2811-M-001-0053. YYK wishes to
thank H.Y. Cheng for helpful discussions and M. Kobayashi for his
encouragement. 

\vskip 2.0cm
\newpage

\centerline{\large\bf APPENDIX A}
\vskip 0.5cm

In this Appendix we supply the details of the Wilson evolution.
The matching conditions at $\mu = M_W$ are given by \cite{Buchalla}
\begin{eqnarray}
C_1(M_W) &=& 0, \nonumber \\
C_2(M_W) &=& 1, \nonumber \\
C_3(M_W) &=& - {\alpha_s(M_W) \over 24 \pi}
E_0(x_t) + {\alpha_{em} \over 6 \pi} {1 \over \sin^2\Theta_W}
[2 B_0(x_t) + C_0(x_t)], \nonumber \\
C_4(M_W) &=& {\alpha_s(M_W) \over 8 \pi} E_0(x_t),
\nonumber \\
C_5(M_W) &=& - {\alpha_s(M_W) \over 24 \pi} E_0(x_t),
\nonumber \\
C_6(M_W) &=& {\alpha_s(M_W) \over 8 \pi} E_0(x_t),
\nonumber \\
C_7(M_W) &=& {\alpha_{em} \over 6 \pi}
\left[4 C_0(x_t) + D_0(x_t) \right],
\nonumber \\
C_8(M_W) &=& 0, \nonumber \\
C_9(M_W) &=& {\alpha_{em} \over 6 \pi}
\left\{4 C_0(x_t) + D_0(x_t)
+ {1 \over \sin^2\Theta_W}
\left[10 B_0(x_t) - 4 C_0(x_t)\right] \right\}, \nonumber \\
C_{10}(M_W) &=& 0, 
\end{eqnarray}
with $x_t = m_t^2/M_W^2$, $m_t$ being the top quark mass.
The functions $B_0, C_0, D_0$ and $E_0$ are the Inami-Lim functions
\cite{inami}:
\begin{eqnarray}
B_0(x) &=& {1 \over 4} \left[{x \over 1-x} +
{x \ln x \over (x - 1)^2} \right]\;, \\
C_0(x) &=& {x \over 8} \left[{x-6 \over x-1} 
+ {3x+2 \over (x - 1)^2} \ln x \right]\;, \\
D_0(x) &=& -{4 \over 9}\ln x - {19x^3 - 25x^2 \over 36 (x-1)^3}
+ {x^2(5x^2 - 2x -6) \over 18(x-1)^4} \ln x\;, \\
E_0(x) &=& -{2 \over 3}\ln x - {x^2(15 -16x + 4x^2) \over 6 (x-1)^4}\ln x
+ {x(18-11x-x^2) \over 12(1-x)^3}\;.
\end{eqnarray}
We adopt the following parameters: $m_t = 170$ GeV, $M_W = 80.2$ GeV,
$\alpha_s(M_W) = 0.118$, $\alpha_{em} = 1/129$, $\sin^2\Theta_{W} = 0.23$
and $\Lambda^{(4)}_{\overline{MS}} = 250$ MeV.

The solution to Eq.~(\ref{RGE-C}) is written as
\begin{equation}
\vec{C}(\mu) = U(t,m_b) \,\, \vec{C}(m_b)\;,
\end{equation}
The evolution function including electroweak penguins is 
\begin{eqnarray}
U(t,m_b,\alpha_{em}) &=& U_f(t,m_b) + {\alpha_{em} \over 4 \pi}
\int_{\ln m_b}^{\ln t} d \ln\mu^{'} U_f(t,\mu^{'})
[\hat{\gamma}_e^{(0)T}]_f U_f(\mu^{'},m_b)
\nonumber \\
&=& U_f(t,M_b) + {\alpha_{em} \over 4 \pi} R_f(t,m_b)\;,
\label{Utmb}
\end{eqnarray}
with
\begin{equation}
U_f(t,m_b) \equiv \exp\left[
\int_{\ln m_b}^{\ln t} d \ln\mu^{'} {\alpha_{s}(\mu^{'}) \over 4 \pi}
[{\hat{\gamma}_s^{(0)T}}]_f \right]\;.
\label{U4}
\end{equation}
For $\mu = m_b = 4.8$ GeV, the values of $C_{i}(m_b)$ are 
\begin{eqnarray}
C_1(m_b) &=& -0.271\;, \hspace{15mm} C_2(m_b) = 1.124\;, \nonumber \\
C_3(m_b) &=& 1.255 \times 10^{-2}\;, \hspace{10mm} 
C_4(m_b) = -2.686 \times 10^{-2}\;, \hspace{10mm}  \nonumber \\
C_5(m_b) &=& 7.805 \times 10^{-3}\;, \hspace{10mm} 
C_6(m_b) = -3.287 \times 10^{-2}\;, \nonumber \\
C_7(m_b) &=& 3.453 \times 10^{-4}\;, \hspace{10mm} 
C_8(m_b) = 3.177 \times 10^{-4}\;, \hspace{10mm} \nonumber \\
C_9(m_b) &=& -9.765 \times 10^{-3}\;, \hspace{10mm} 
C_{10}(m_b) = 2.240 \times 10^{-3}\;.
\end{eqnarray}
Values of the Wilson coefficients at different energy scales $\mu=$ 1.0
GeV, 1.5 GeV, 2.0 GeV, 2.5 GeV, 3.0 GeV and 4.8 GeV are listed in
Table III.

\vskip 1.0cm
\centerline{\large\bf APPENDIX B}
\vskip 0.5cm

Below we summarize the parameters we have adopted in the numerical
analysis of this work:
\begin{enumerate}
\item Masses, decay constants and lifetimes:
\begin{center}
$M_{{\pi}^{\pm}} =0.139$ GeV, \hspace{20mm} $M_{{\pi}^{0}} = 0.135$ GeV,\\
$M_{K} =0.49$ GeV, \hspace{20mm} $M_{B} = 5.28$ GeV, \\
$m_u = 4.5$ MeV, \hspace{20mm} $m_d = 8.1$ MeV, \\
$m_s = 100$ MeV, \hspace{20mm} $m_c = 1.5$ GeV, \\
$m_b = 4.8$ GeV, \hspace{20mm} $m_t = 170$ GeV,\\
$M_W = 80.2$ GeV,\hspace{20mm} $f_{B} = 190$ MeV,\\
$f_{\pi} = 130$ MeV, \hspace{20mm} $f_{K} = 160$ MeV,\\
$\tau_{B^0}=1.55$ ps, \hspace{20mm} $\tau_{B^-}=1.65$ ps.
\end{center}

\item QCD and electroweak parameters:
\begin{center}
$G_F = 1.16639 \times 10^{-5}$ GeV$^{-2}$, 
\hspace{20mm}$\Lambda_{\overline{MS}}^{(4)} = 250$ MeV,\\  
$\alpha_s(M_Z) = 0.117$,\hspace{20mm} $\alpha_{em} = 1/129$, \\
$\lambda = 0.2196$, \hspace{20mm} $A = 0.819$, \\
$R_b = \sqrt{\rho^2 + \eta^2} = 0.38$. \\
\end{center}

\item Meson wave functions:
\begin{eqnarray}
\phi_B(x) &=& N_B
x^2 (1-x)^2\exp\left[-\frac{1}{2}\left(\frac{xM_B}{\omega_B}\right)^2
-\frac{\omega_B^2 b^2}{2}\right]\;,
\nonumber \\
N_B&=&91.7835\;\;{\rm GeV}\;,\;\;\;\;\omega_B = 0.4 \,\,{\rm GeV}\;,
\nonumber\\
\phi_{\pi}(x) &=& \frac{3}{\sqrt{2N_c}}f_\pi x(1-x)
[1 + 0.8(5(1-2x)^2-1)]\;,
\nonumber \\
\phi^{'}_{\pi}(x) &=&\frac{3}{\sqrt{2N_c}}f_\pi x(1-x)\;,
\nonumber\\
\phi_{K}(x) &=& \frac{3}{\sqrt{2N_c}}f_K x(1-x)[1 + 0.51(1-2x)
+ 0.3(5(1-2x)^2-1)]\;,
\nonumber \\
\phi^{'}_{K}(x) &=& \frac{3}{\sqrt{2N_c}}f_K x(1-x)\;.
\nonumber
\end{eqnarray}

\end{enumerate}

\newpage

{\bf \Large Figure Captions}
\vspace{10mm}

\begin{enumerate}
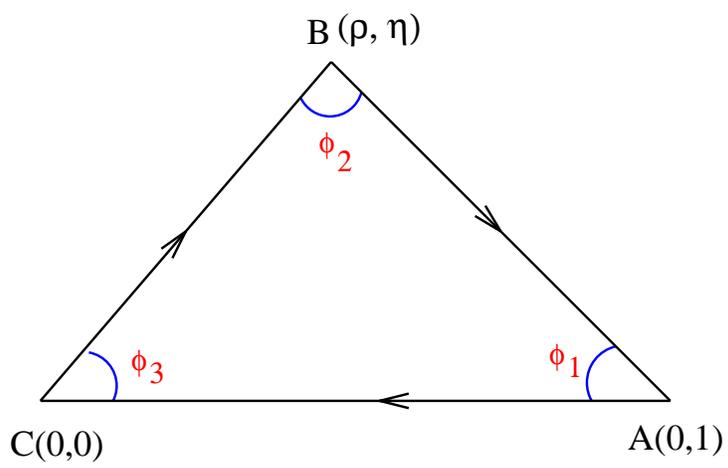
\item Fig. 1: Unitarity triangle and the definition of the angles
$\phi_{i}$.

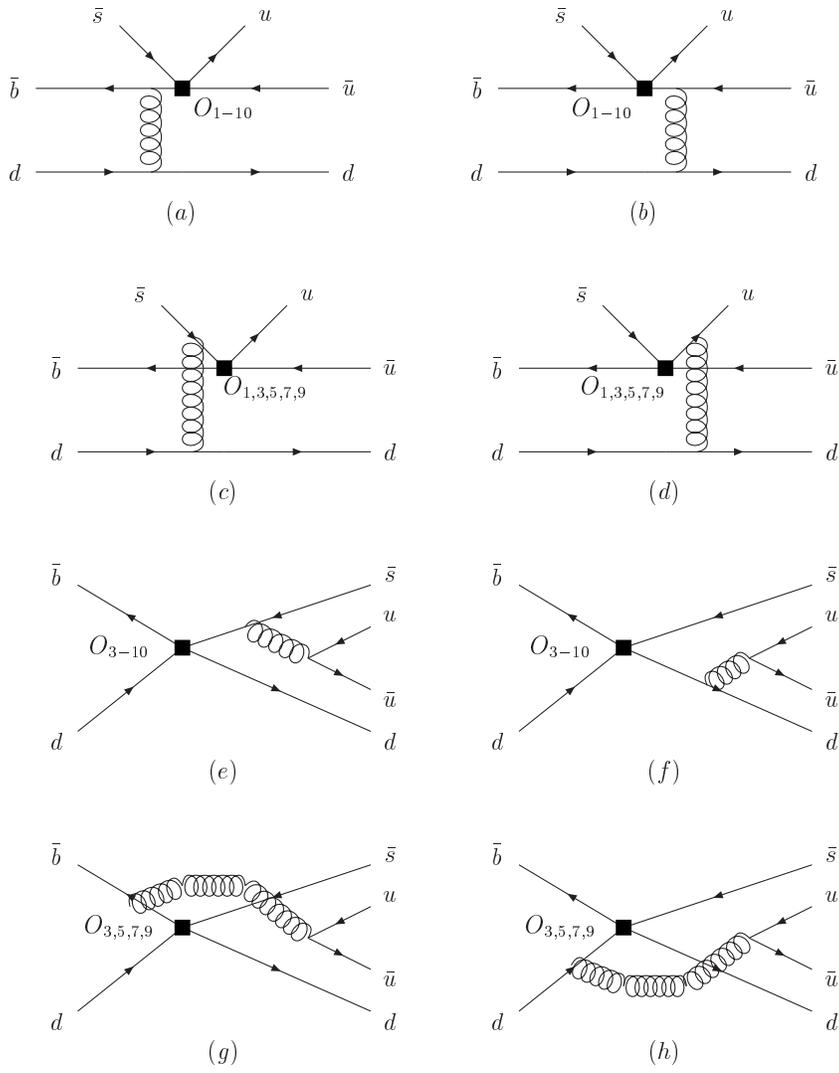
\item Fig. 2: Feynman diagrams for the
$B^{\pm}\to \bar{K}^{0}\pi^{\pm}$ decays.

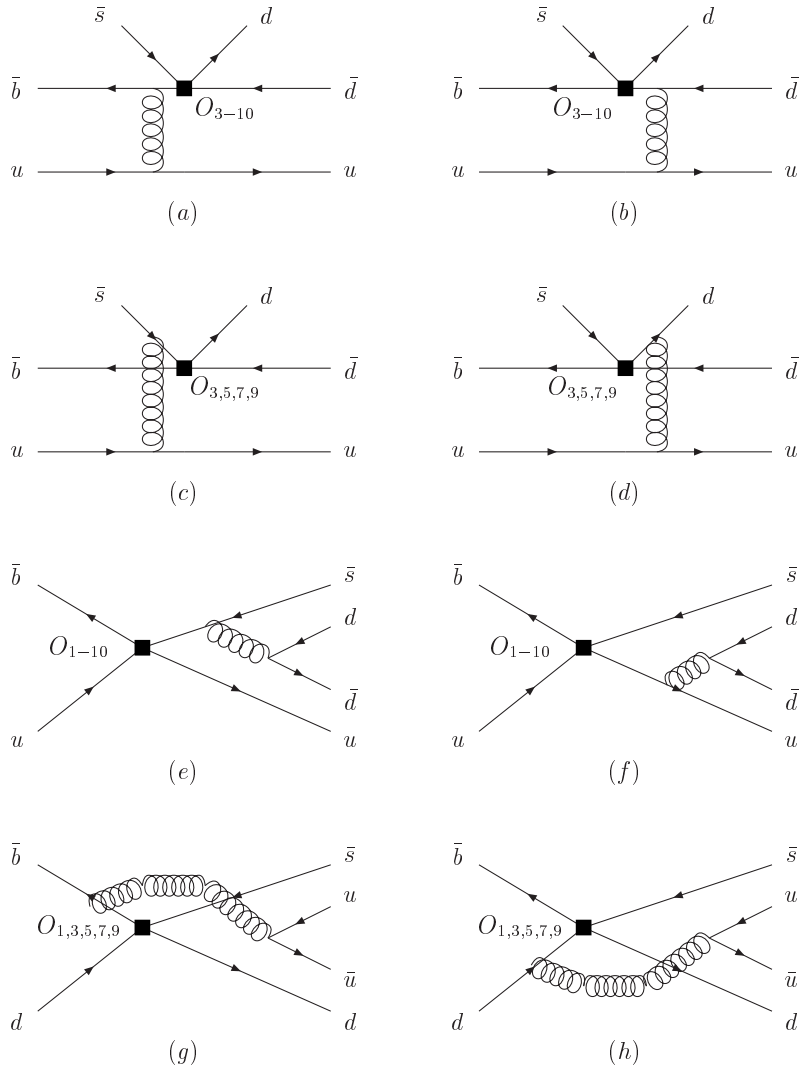
\item Fig. 3: Feynman diagrams for the $B_d^{0} \to K^{\pm}\pi^{\mp}$
decays.

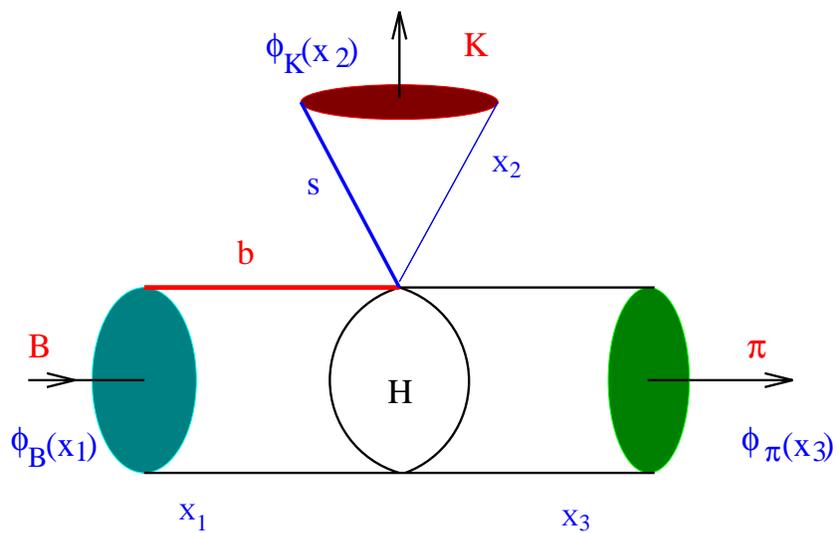
\item Fig. 4: Factorization of the $B \to K \pi $ decays in the PQCD
approach.

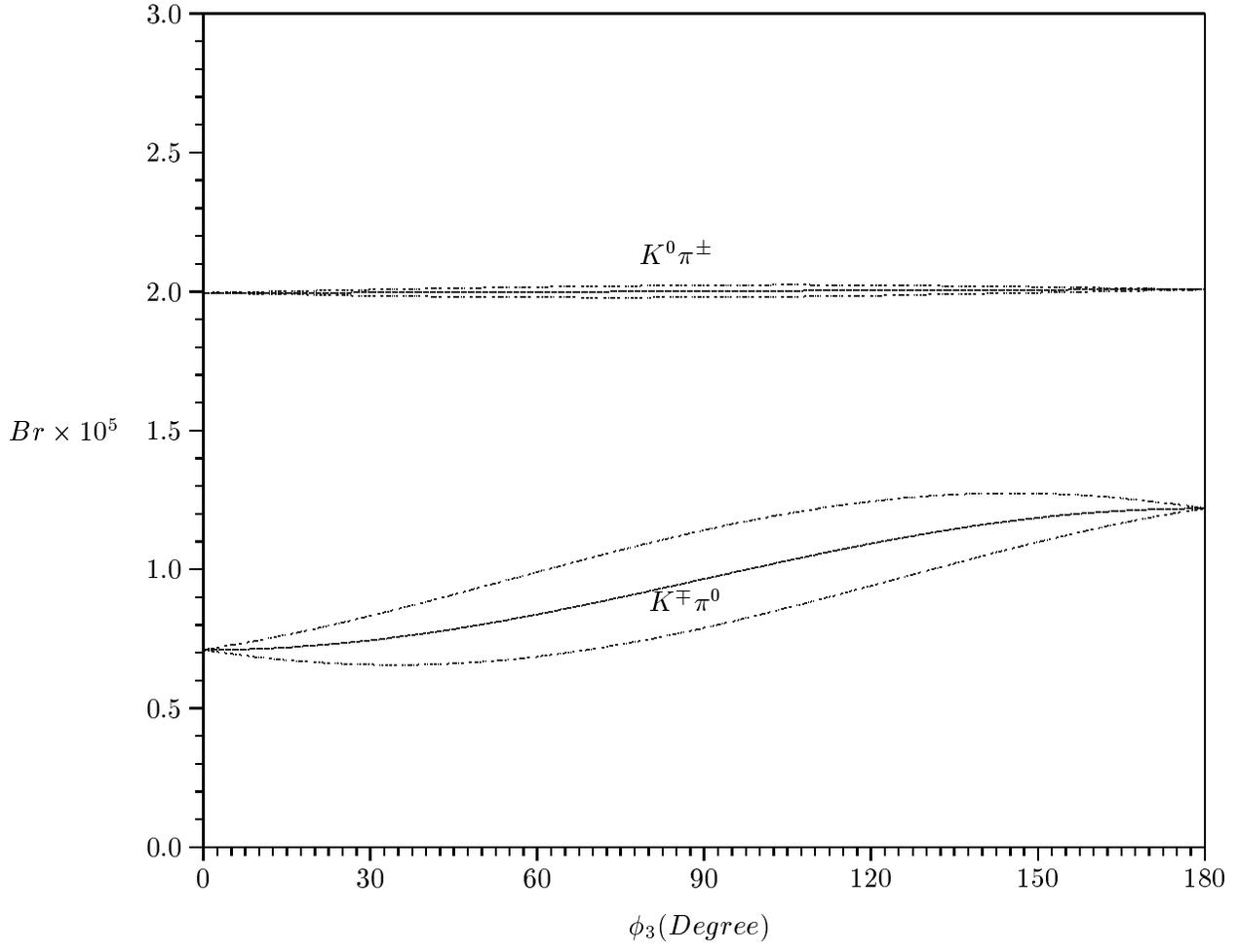
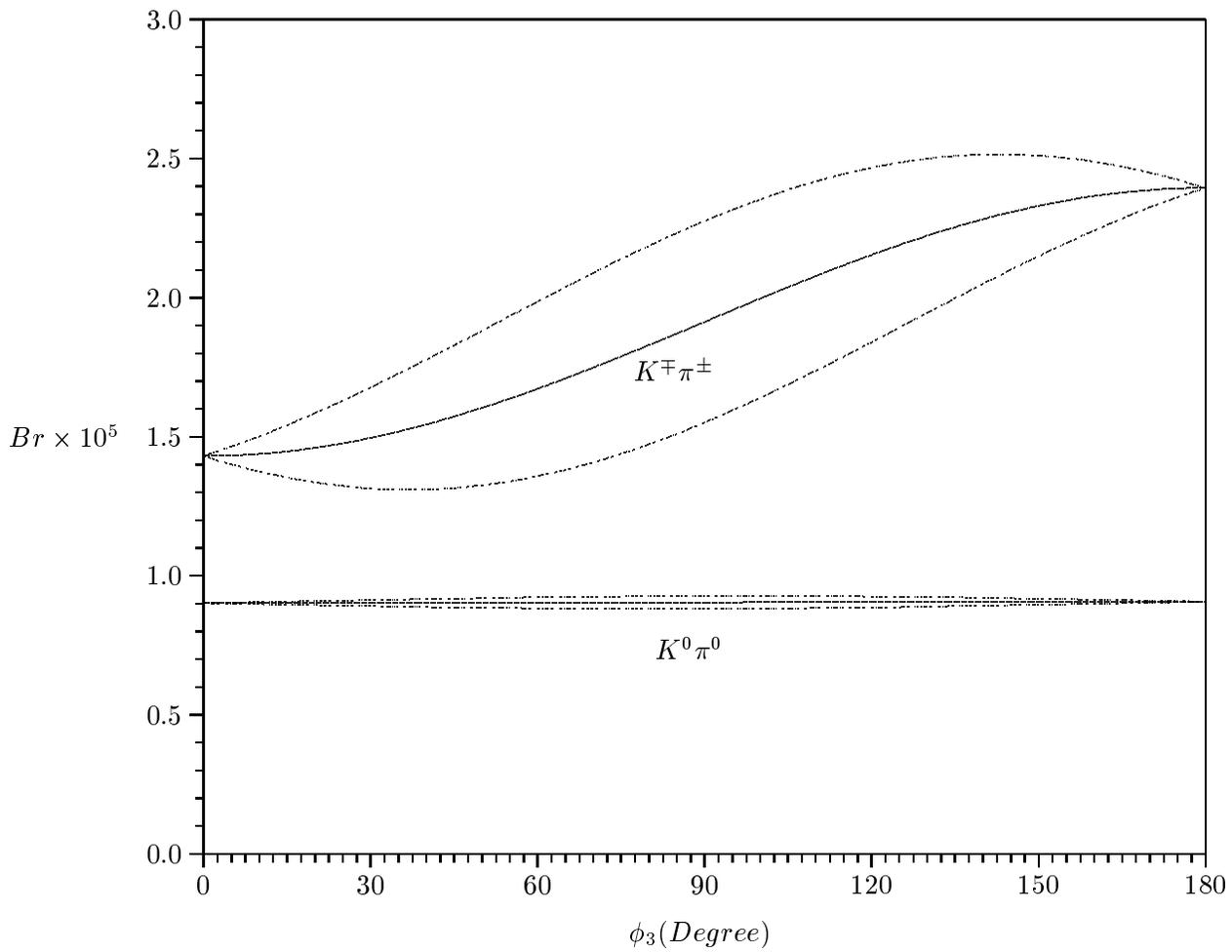
\item Fig. 5: Dependence of the branching ratios of the $B \to K\pi$
decays on $\phi_3$ with the upper (lower) dashed line corresponding to
the ${\bar B}$ ($B$) meson decays.

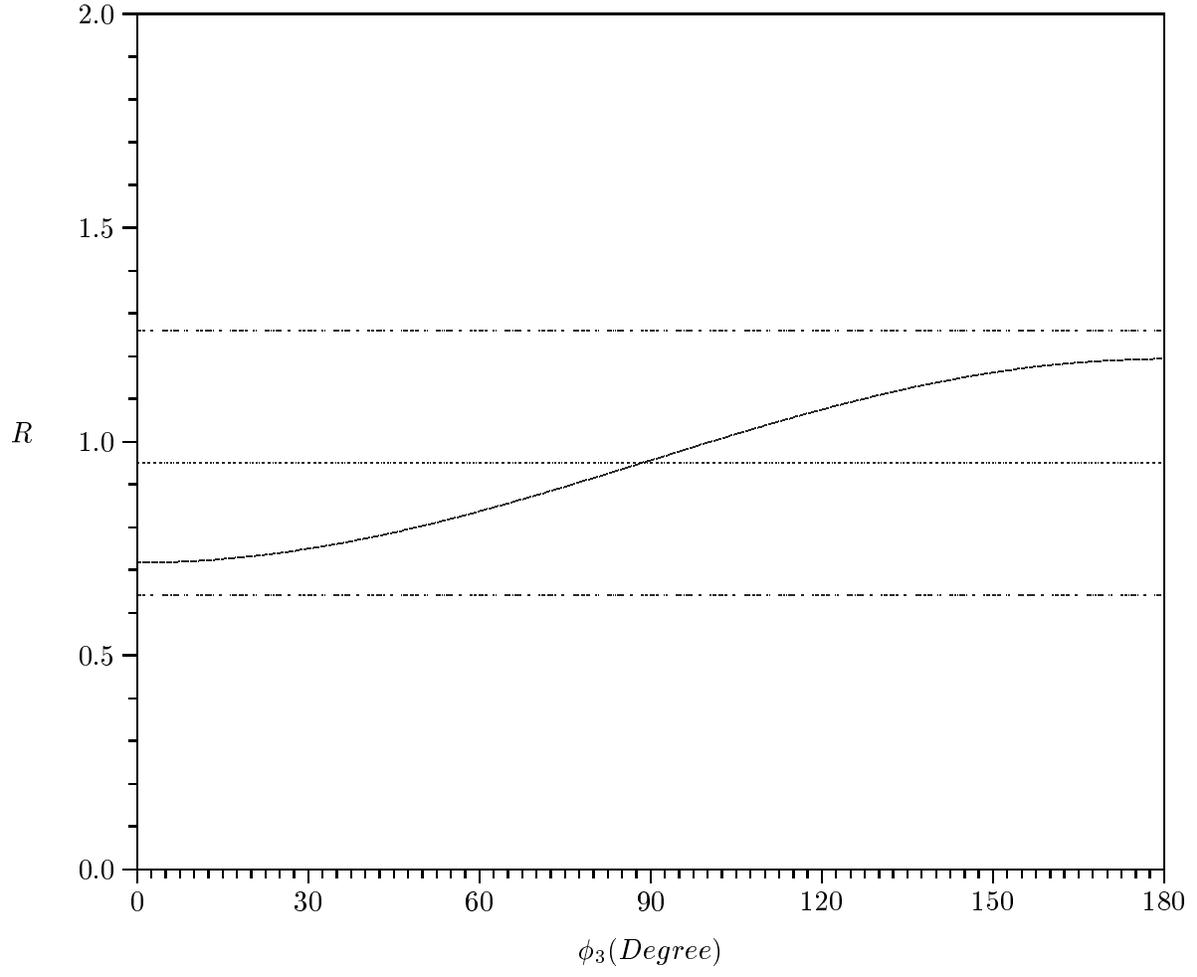
\item Fig. 6: Dependence of the ratio $R$ on $\phi_3$.
The dashed (dotted) lines correspond to the bounds (central value)
of the data.

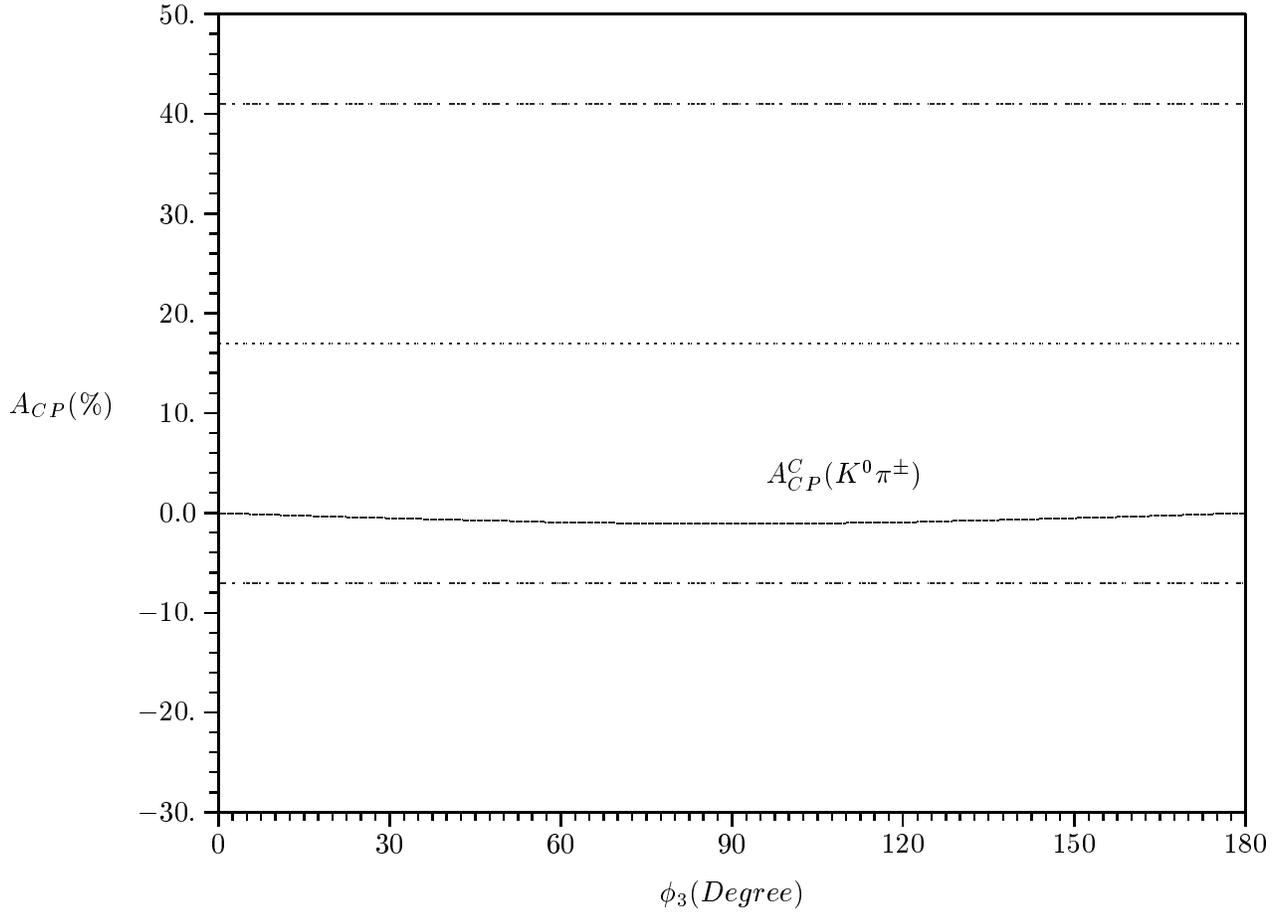
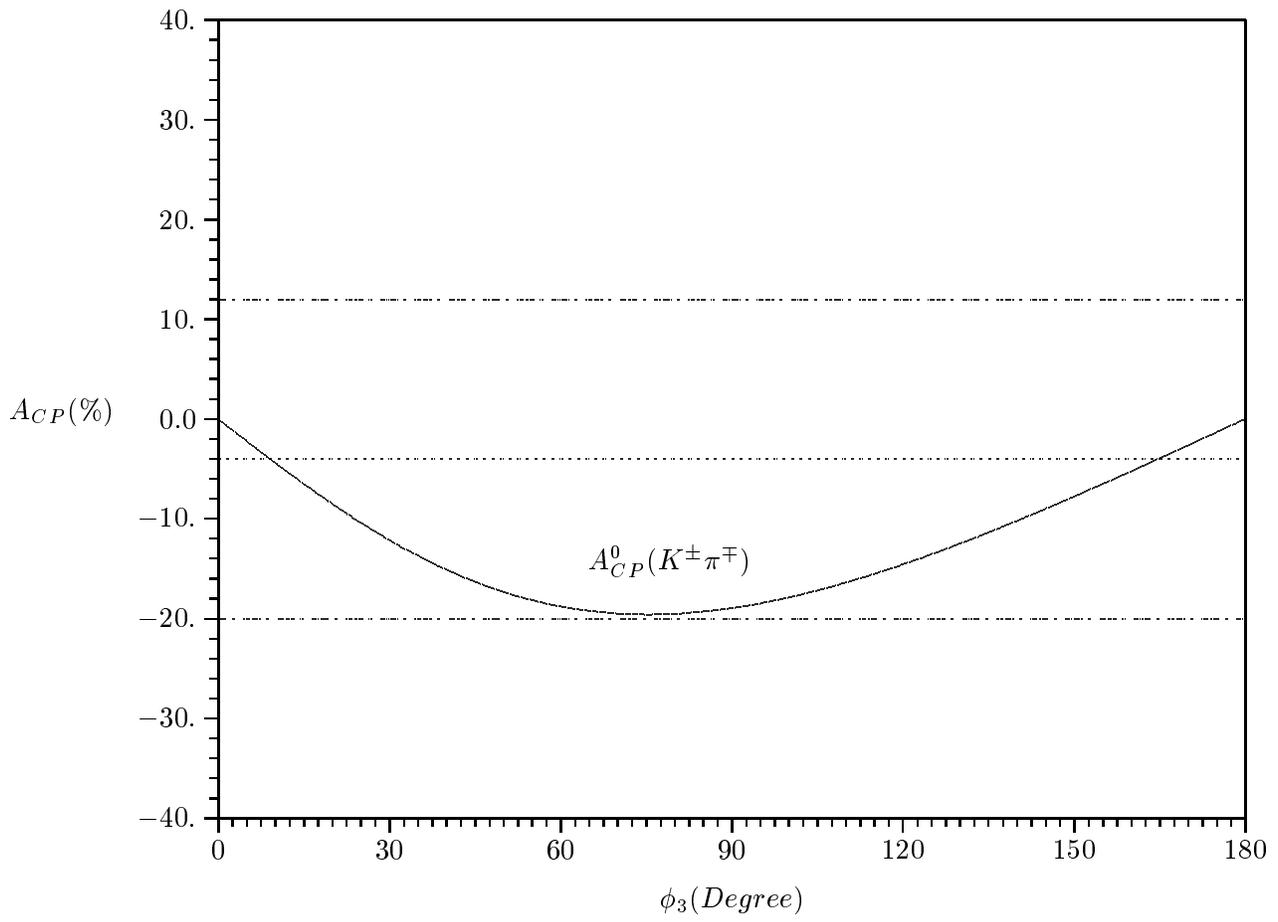
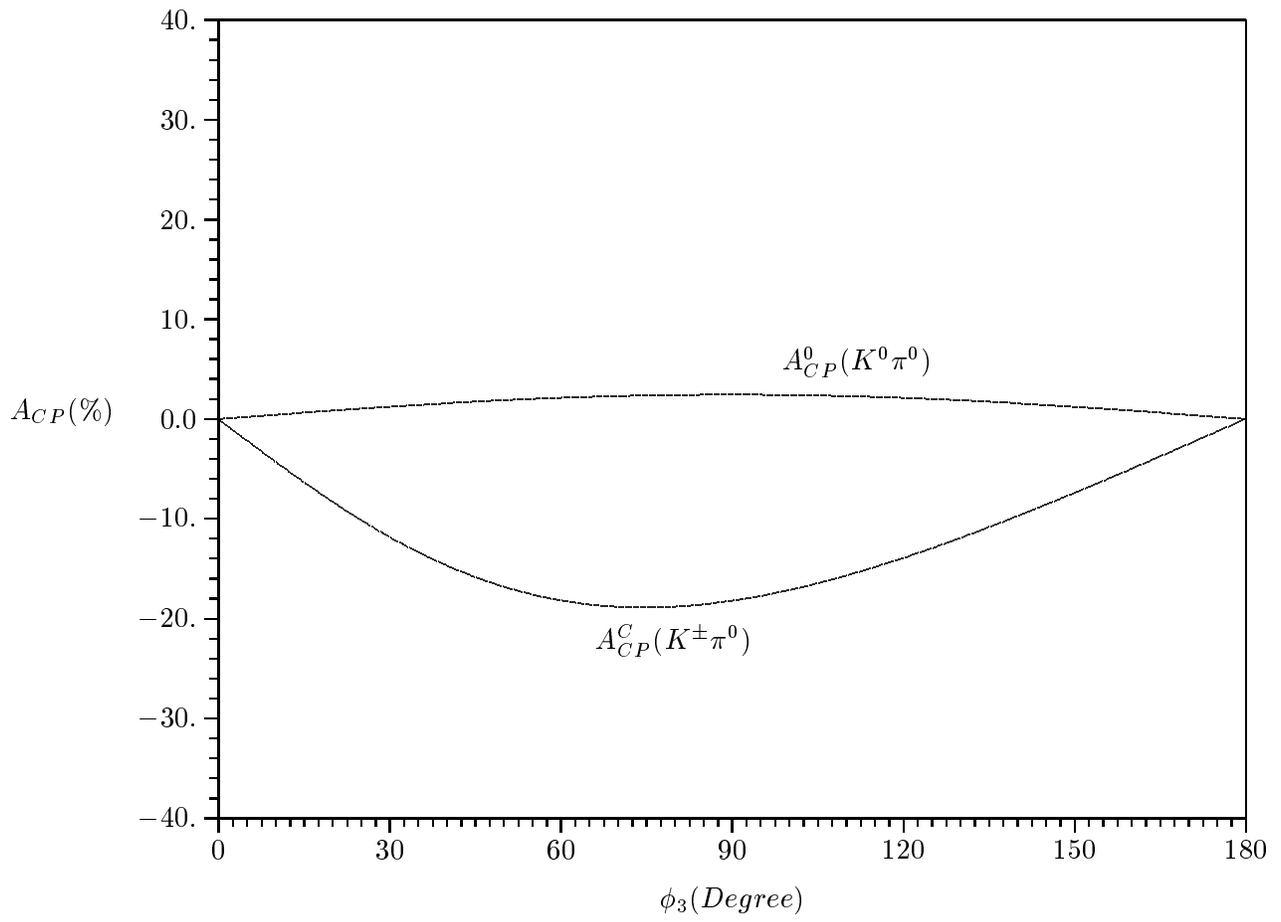
\item Fig. 7: Dependence of CP asymmetries on $\phi_3$.
The dashed (dotted) lines correspond to the bounds (central value)
of the data of the $B^{\pm}\to K^0\pi^{\pm}$ decays in (a)  and
the $B_d^0 \to K^{\pm} \pi^{\mp}$ in (b).

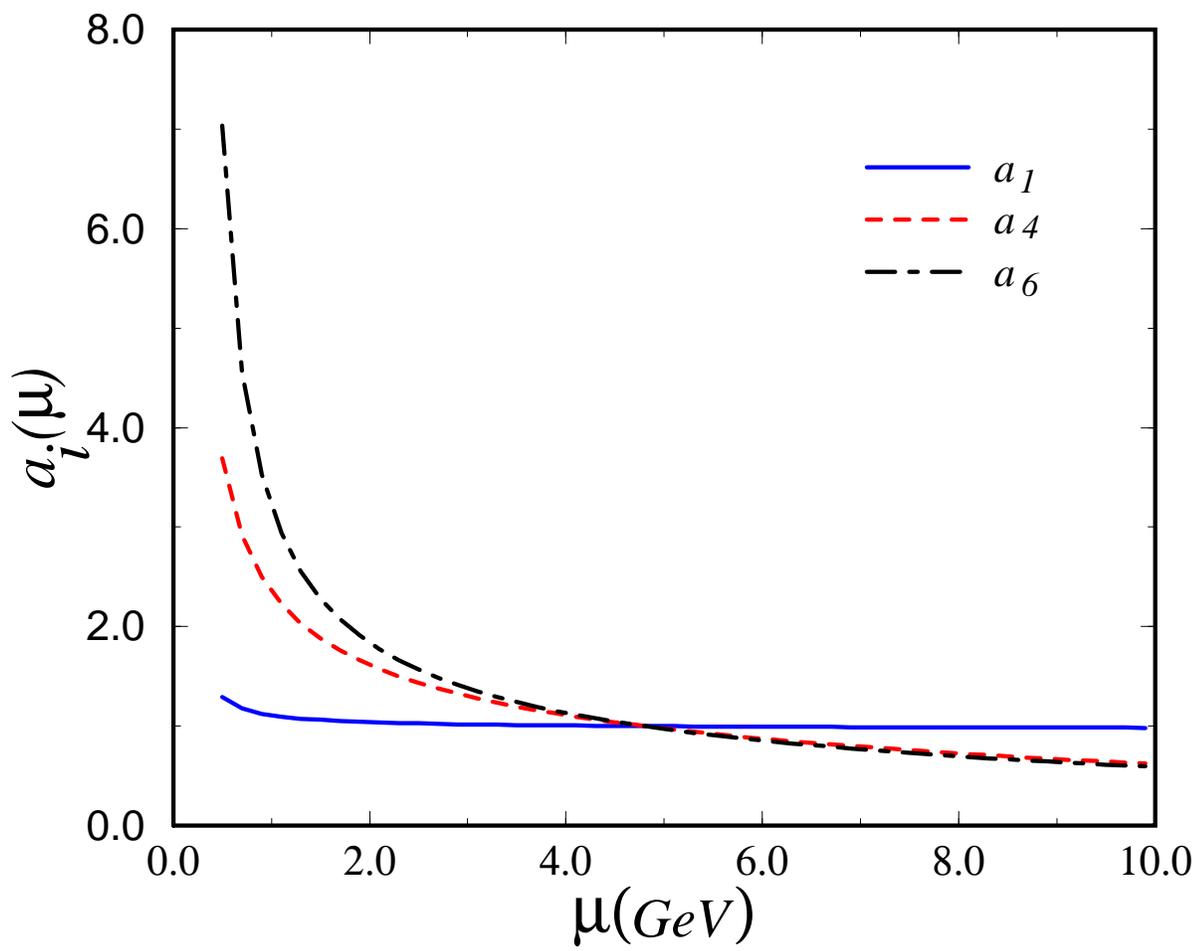
\item Fig. 8: RG evolution of the Wilson coefficients
$a_{i}(\mu)$, $i = 1$ ,4 ,6, normalized by their values at $\mu = m_b$.

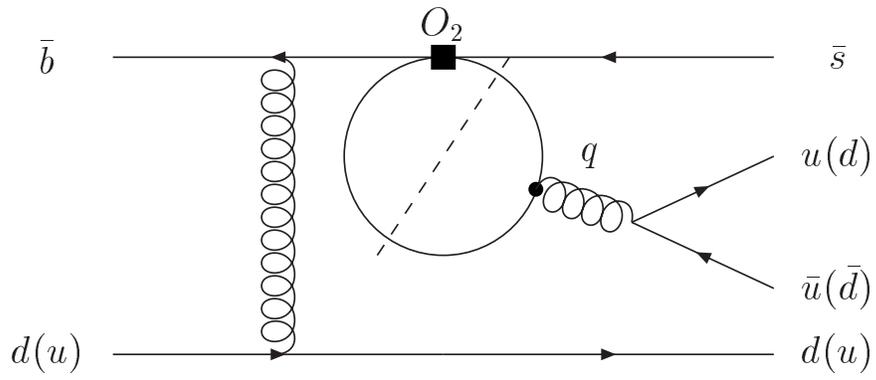
\item Fig. 9: Frcation contribution to $F_{e6}^P$  as a
function of $\alpha_s(t)/\pi$.

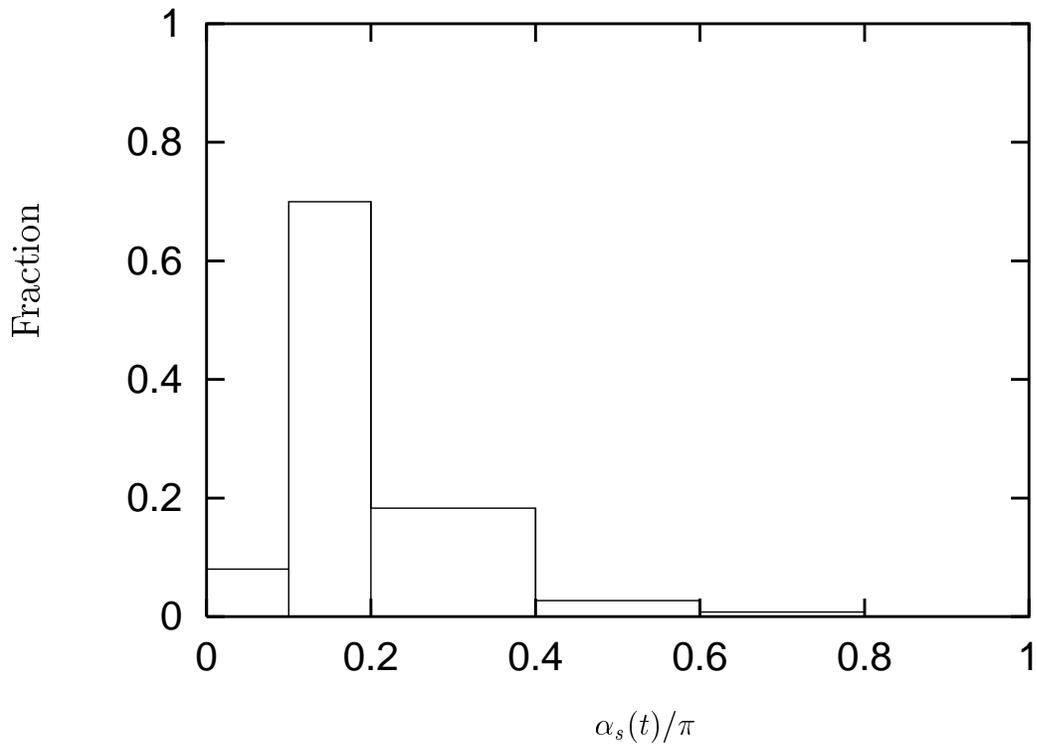
\item Fig. 10: Feynman diagram for an induced $c$ ($u$)-quark loop.
\end{enumerate}

\newpage

\begin{table}[ht]
\vspace*{0.5cm}
\begin{center}
\begin{tabular}{||c||c||}   \hline \hline
$F_e$ &  $5.577 \times 10^{-1}$    \\
$F_e^P$ & $-5.537 \times 10^{-2}$   \\
$F_a^P$ & $3.333\times 10^{-3} + \,\, i \,\, 3.181\times 10^{-2}$ \\
$M_e$ & $-0.942\times 10^{-3} + \,\, i \,\, 
3.385\times 10^{-3}$  \\      
$M_e^P$ & $2.931\times 10^{-5} - \,\, i \,\, 
1.304\times 10^{-4}$   \\       
$M_a^P$ & $-9.397\times 10^{-5} - \,\, i \,\, 
1.918\times 10^{-4}$   \\       
\hline \hline
\end{tabular}
\end{center}
\caption{Contribution to the $B^0 \to K^+ \pi^-$ decay from
each form factor and nonfactorizable amplitude.
\label{table3}}
\end{table}

\newpage

\begin{table}[ht]
\vspace*{0.5cm}
\begin{center}
\begin{tabular}{||c||c|c||}   \hline \hline
$q^2$& $Re[G]$ & $Im[G]$ \\
\hline \hline
$m_B^2$ &  -0.760 & 2.025          \\
$m_B^2/2$ & 0.139 & 1.775           \\
$m_B^2/3$ & 0.912 & 1.178           \\
$m_B^2/3.5$ & 1.288 & 0.392            \\
$m_B^2/4$ & 1.162 &  0.000        \\
\hline \hline
\end{tabular}
\end{center}
\caption{Real and imaginary parts of the charm quark loop
contribution $G(q^2)=-4 \int du u (1-u)
\ln [m_c^2 -q^2 u(1-u))]$ in the BSS mechanism.
\label{table2}}
\end{table}
\newpage

\begin{table}[ht]
\vspace*{0.5cm}
\begin{center}
\begin{tabular}{||c||c|c|c|c|c|c||}   \hline
 &\multicolumn{6}{p{20em}||}
{ \hspace{39mm} $\Lambda^{(4)}_{\overline{MS}} = 250 \,\, MeV$} \\
\hline \hline
$\mu$ &  1.0 GeV & 1.5 GeV & 2.0 GeV & 2.5 GeV & 3.0 GeV & 4.8 GeV \\
\hline \hline
$\alpha_s(\mu)$ & 0.5439 & 0.4208 
& 0.3626 & 0.3275 & 0.3034 & 0.2552  \\
\hline \hline
$C_1$ &-0.650 & -0.510 & -0.435 & -0.385 & -0.349 & -0.271 \\
$C_2$ & 1.362 & 1.268 & 1.219 &  1.189  & 1.168  & 1.124  \\
\hline
$C_3$ & 0.036 & 0.027 & 0.022 &  0.019  & 0.017 & 0.013  \\
$C_4$ &-0.063 & -0.050 & -0.043 & -0.038 & -0.035  & -0.027 \\
$C_5$ & 0.015 &  0.013 & 0.012   & 0.011  & 0.010  & 0.008  \\
$C_6$ &-0.102 & -0.074 & -0.060 &  -0.051  & -0.045 & -0.033  \\
\hline
$C_7/\alpha_{em}$ & 0.040 & 0.035 & 0.035 & 0.036 & 0.038  & 0.045  \\
$C_8/\alpha_{em}$ & 0.128 & 0.091 & 0.073 & 0.062  & 0.055 & 0.041  \\
$C_9/\alpha_{em}$ & -1.509 & -1.416 & -1.366 & -1.334 & -1.311 & -1.260 \\
$C_{10}/\alpha_{em}$ & 0.695 & 0.546 & 0.465 & 0.412 & 0.373 & 0.289  \\
\hline \hline
\end{tabular}
\end{center}
\caption{Values of the running coupling constant $\alpha_s$
and the Wilson coefficients $C_{i}$ with
$\Lambda^{(4)}_{\overline{MS}} = 250$ MeV for different energy scales
$\mu = 1.0$, 1.5, 2.0, 2.5, 3.0, and 4.8 GeV.
\label{table1}}
\end{table}

\end{document}

%% file: sanda.tex
\def\be{\begin{equation}}
\def\ee{\end{equation}}
\def\ba{\begin{eqnarray}}
\def\ea{\end{eqnarray}}

\def\to{\rightarrow}

\def\sk{\vskip 1cm}

\def\PR#1#2#3 {{\it Phys. Rev. }{\bf D#1} #2 {(#3)}}
\def\PRL#1#2#3 {{\it Phys. Rev. Lett. }{\bf #1} #2 {(#3)}}
\def\PL#1#2#3 {{\it Phys. Lett. }{\bf #1} #2 {(#3)}}
\def\AP#1#2#3 {{\it Ann, Phys. }{\bf #1} #2 {(#3)}}
\def\ZP#1#2#3 {{\it Z. Phys. }{\bf #1} #2 {(#3)}}
\def\NP#1#2#3 {{\it Nucl. Phys. }{\bf #1} #2 {(#3)}}
\def\MPL#1#2#3 {{\it Mod. Phys. Lett. }{\bf #1} #2 {(#3)}}
\def\NC#1#2#3 {{\it Nuov. Cim. }{\bf #1} #2 {(#3)}}
\def\PREP#1#2#3 {{\it Phys. Report }{\bf #1} #2 {(#3)}}
\def\PROG#1#2#3 {{\it Prog. Theor. Phys. }{\bf #1} #2 {(#3)}}

\def\sq2{{1\over{\sqrt{2}}}}

\def\dbar{\overline d}

\def\sbar{\overline s}
\def\bbar{\overline B}

\def\g5{\gamma_5}

\renewcommand{\Im}{\mbox{Im}\,}